\documentclass[12pt]{article}
\usepackage{amsmath}
\usepackage{graphicx,psfrag,epsf}
\usepackage{enumerate}
\usepackage[left=1in,right=1in,top=1in,bottom=1in]{geometry}

\usepackage{setspace}
\usepackage{graphicx}
\usepackage{wrapfig}
\usepackage{epstopdf}
\usepackage{subcaption}
\usepackage{caption}
\usepackage{multirow}
\usepackage{nicefrac}
\usepackage{booktabs}
\usepackage{amsmath}
\usepackage{amsthm}
\usepackage[algo2e]{algorithm2e} 
\usepackage{verbatim} 
\usepackage{amssymb}
\usepackage{bbm}
\usepackage{mathrsfs}
\usepackage[english]{babel}
\usepackage{graphicx,epstopdf}
\RequirePackage[authoryear]{natbib}
\usepackage{hyperref}
\hypersetup{colorlinks=true,citecolor=blue}
\hypersetup{linkcolor=blue}
\usepackage{algorithm}
\usepackage[noend]{algpseudocode}
\usepackage{tikz}

\newcommand{\be}{\begin{eqnarray}}
\newcommand{\ee}{\end{eqnarray}}
\newcommand{\ba}{\begin{eqnarray*}}
\newcommand{\ea}{\end{eqnarray*}}

\newcommand{\bu}{\boldsymbol{u}}
\newcommand{\bv}{\boldsymbol{v}}

\newcommand{\bx}{\boldsymbol{x}}

\usepackage{bm}

\usepackage{array}
\newcolumntype{L}[1]{>{\raggedright\let\newline\\\arraybackslash\hspace{0pt}}m{#1}}
\newcolumntype{C}[1]{>{\centering\let\newline\\\arraybackslash\hspace{0pt}}m{#1}}
\newcolumntype{R}[1]{>{\raggedleft\let\newline\\\arraybackslash\hspace{0pt}}m{#1}}

\usepackage{array}

\include{math-commands}

\newcommand{\bbeta}{\boldsymbol{\beta}}
\title{BOPIM: Bayesian Optimization for influence maximization on temporal networks}

\author{%
  Eric Yanchenko\footnote{Akita International University, Global Connectivity Program, Akita, Japan}
  }

\begin{document}

\doublespacing

\begin{titlepage}

\maketitle
\thispagestyle{empty}
\begin{abstract}
\noindent
The goal of influence maximization (IM) is to select a small set of seed nodes which maximizes the spread of influence on a network. In this work, we propose $\mathsf{BOPIM}$, a Bayesian Optimization (BO) algorithm for IM on temporal networks. The IM task is well-suited for a BO solution due to its expensive and complicated objective function. There are at least two key challenges, however, that must be overcome, primarily due to the inputs coming from a cardinality-constrained, non-Euclidean, combinatorial space. The first is constructing the kernel function for the Gaussian Process regression. We propose two kernels, one based on the Hamming distance between seed sets and the other leveraging the Jaccard coefficient between node's neighbors. The second challenge is the acquisition function. For this, we use the Expected Improvement function, suitably adjusting for noise in the observations, and optimize it using a greedy algorithm to account for the cardinality constraint. In numerical experiments on real-world networks, we prove that $\mathsf{BOPIM}$ outperforms competing methods and yields comparable influence spreads to a gold-standard greedy algorithm while being as much as ten times faster. In addition, we find that the Hamming kernel performs favorably compared to the Jaccard kernel in nearly all settings, a somewhat surprising result as the former does not explicitly account for the graph structure. Finally, we demonstrate two ways that the proposed method can quantify uncertainty in optimal seed sets. To our knowledge, this is the first attempt to look at uncertainty in the seed sets for IM.

\end{abstract}


\end{titlepage}

\section{Introduction}\label{sec:intro}

Influence maximization (IM) is a canonical network science problem where the goal is to select a small set of seed nodes in order to maximize the influence spread on a graph \citep{kempe2003maximizing, chen2009efficient}. Over the past two decades, IM has garnered significant research interest due to its relevance in online marketing campaigns \citep{domingos2001mining, bharathi2007competitive, chen2010scalable}, rumor spreading \citep{kempe2003maximizing, murata2018extended}, public health campaigns \citep{wilder2017uncharted, yadav2017influence, wilder2018end, yadav2018bridging}, vaccine strategies \citep{holme2004efficient, holme_vacc}, and more. Traditionally, the problem assumes the graph to be static, i.e., nodes and edges are fixed. In many situations, however, this may be unrealistic. For example, consider a marketing campaign on X where a company promotes a new product. As users and followers change daily, an effective IM algorithm must account for this dynamism. Therefore, recent work has looked at the IM problem on temporal networks \citep{holme2012temporal} where edges change with time. Since this combinatorial optimization problem is NP-hard, heuristics must be employed. Some methods use a greedy algorithm and probabilistic approximation of the expected influence spread \citep{aggarwal2012influential, osawa2015selecting, erkol2020influence}, while others eschew costly influence spread calculations by simply ranking nodes in order of importance \citep{michalski2014seed, murata2018extended, michalski2020entropy}. We refer the interested reader to \cite{yanchenko2024influence} for a review of temporal IM.

While IM is a well-known problem in the network, information and computer sciences, it has received relatively little attention from statisticians. This has led to several consequences on IM methodology development, not least of which being a lack of uncertainty quantification. Additionally, as the IM task is an optimization problem, statistics has a whole suit of techniques to offer that hitherto have been unexplored. One such approach is {\it Bayesian Optimization} (BO) which has emerged as a popular paradigm for optimizing complicated objective functions \citep{snoek2012practical, shahriari2015taking, frazier2018tutorial}. Machine learning hyper-parameter tuning \citep{snoek2012practical, wu2019hyperparameter, turner2021bayesian}, experimental design \citep{frazier2015bayesian, shahriari2015taking, greenhill2020bayesian} and computer experiments \citep{surer2023sequential, sauer2023active} are just a few of the many applications where BO has been successfully implemented. BO approximates a complicated objective function with a simple surrogate function and fits this model via Bayesian regression. The statistical model is maximized with an acquisition function before evaluating this new point on the original objective function. Finally, this point is added to the data set and the process repeats.

In this work, we propose $\mathsf{BOPIM}$, Bayesian OPtimization for Influence Maximization, a BO algorithm for IM on temporal networks. The key challenges in applying a BO framework to our problem are in constructing the kernel function and optimizing the acquisition function. To address these, we propose two kernels, one based on the Hamming distance between seed sets, and the other on the similarity in node's neighbors. We then leverage the Expected Improvement acquisition function to select the next candidate point, appropriately accounting for the noise in our observations. We optimize this function using a greedy algorithm which shares many similarities to the Trust Region framework \citep[e.g.,][]{eriksson2019scalable, wan2021think, papenmeier2023bounce}, but also accounts for the cardinality constraint. In numerical experiments on real-world networks, $\mathsf{BOPIM}$ yields seed sets with influence spreads comparable to those of a ``gold-standard'' greedy algorithm but at a fraction of the computational time.

There are two primary contributions of this work. First, to the knowledge of the author, this is that first time that a BO framework has been applied to the IM problem. The key challenges stem from the inputs having a cardinality constraint, while also corresponding to non-Euclidean space, i.e., nodes on a graph. There have been several contributions to the combinatorial BO literature, including \cite{garnett2010bayesian, baptista2018bayesian, oh2019combo, ru2020bayesian, buathong2020kernels, daulton2022bayesian, papenmeier2023bounce}. The proposed method is inspired by these papers, most notably in the construction of the kernel, and optimization of the acquisition function. The cardinality constraint, however, means that these methods cannot easily be applied directly. For example, we show in detail why $\mathsf{COMBO}$ \citep{oh2019combo} cannot be used with such a constraint. Thus, this work addresses the various challenges of adapting the BO framework to the IM problem.

Second, the BO algorithm naturally provides measures of uncertainty on the optimal seed sets. While there are many methods developed for IM, there has been little emphasis on the uncertainty in the results. The proposed method not only provides uncertainty quantification of the total influence spread on the network, but also a richer understanding of a node's importance in the optimal seed set. Indeed, $\mathsf{BOPIM}$ can inform us if there are many seed sets which yield similar influence spreads, and/or how confident we are that a node should be in the optimal seed set, as opposed to a simple binary result on its inclusion/exclusion. We discuss two different ways that $\mathsf{BOPIM}$ can help answer these questions.

The layout of the paper is as follows. For the remainder of Section \ref{sec:intro}, we introduce notation and give a formal problem statement. Section \ref{sec:bo} presents the main algorithm while Section \ref{sec:exp} is devoted to numerical experiments. In Section \ref{sec:uq} we discuss uncertainty quantification, and share concluding thoughts and future directions in Section \ref{sec:conc}.

\subsection{Notation and problem statement}
Let $\mathcal G=(G_1,\dots,G_T)$ be a temporal graph with $T$ snapshots or ``slices,'' where a graph snapshot $G_t=(V,E_t)$ is defined by node set $V$ and edge set $E_t$. We let $|V|=n$ be the number of nodes and define $m=|\cup_{t=1}^T E_t|$ as the number of unique edges. Given an influence diffusion process and integer $k<n$, the goal of IM is to find a set of nodes $S$ with $|S|=k$ such that if the nodes in $S$ are ``influenced'' at time $t=1$, then the spread of influence at time $T+1$ is maximized. 

We first must choose an influence diffusion mechanism, though the proposed method can easily be adapted to any diffusion mechanism. Some of the most popular choices are the Independent Cascade (IC) \citep{wang2012scalable, wen2017online} and Linear Threshold \citep{chen2010scalableaa, goyal2011celf++} models. We adopt the Susceptible-Infected (SI) model from epidemiological studies \citep{osawa2015selecting, murata2018extended}. In the SI model, a node is either in a susceptible ($\mathscr{S}$) or infected ($\mathscr{I}$) state. A node is infected if it was influenced at some point during the process, and susceptible otherwise. At time $t=1$, nodes in the seed set are set to state $\mathscr{I}$, and all other nodes initialize to state $\mathscr{S}$. At time $t$, a node in state $\mathscr{I}$ attempts to influence all of its neighbors. Specifically, if node $i$ is in state $\mathscr{I}$, node $j$ is in state $\mathscr{S}$, and there exists an edge between node $i$ and $j$ at time $t$, then node $j$ will be in state $\mathscr{I}$ at time $t+1$ with probability $\lambda$. The process concludes at time $T+1$ and the number of nodes in state $\mathscr{I}$ corresponds to the total influence spread. If we let $\sigma(S)$ be the expected number of influenced nodes at time $T+1$ for seed set $S$, the IM problem seeks the set of nodes of size $k$ that maximizes $\sigma(S)$ on $\mathcal G$, i.e.,
\begin{equation}\label{eq:prob}
    S^*
    =\arg\max_{S\subseteq V,|S|=k}\sigma(S).
\end{equation}
Finding $S^*$ is a combinatorial optimization problem with a cardinality constraint and has been shown to be NP-hard under many popular diffusion mechanisms \citep{li2018influence} so finding the global optimum by brute-force calculation is infeasible even for moderate $n$. Indeed, as the influence spread function is typically computed using Monte Carlo (MC) simulations, even evaluating $\sigma(S)$ is costly. Many IM algorithms efficiently compute the influence spread function and apply a greedy algorithm to choose the seed nodes, while others skip evaluation of the objective function and instead use a heuristic to rank nodes by influence. In the remainder of this work, we describe a BO algorithm to approximate the solution in (\ref{eq:prob}).

\section{Bayesian Optimization}\label{sec:bo}
In this section, we present the main contribution of this paper, the $\mathsf{BOPIM}$ algorithm, particularly emphasizing the unique challenges posed by the temporal IM problem.

\subsection{Objective function}\label{sec:obj}
The first step in a BO algorithm is the initial evaluation of the objective function. Let $f:\mathcal X\to\mathbb R$ be some real-valued function defined on domain $\mathcal X$ that we wish to maximize. BO is particularly useful when $f(\cdot)$ is computationally expensive to evaluate, lacks special structure, e.g., concavity, and/or has unavailable first and second order derivatives. For the temporal IM problem, let $\bx\in\mathcal X=\{0,1\}^n$ be such that $x_j=1$ if node $j$ is in the seed set, and 0 otherwise, for $j=1,\dots, n$. Since the seed set is limited to $k$ nodes, we impose the cardinality constraint that $\sum_{j=1}^n x_j=k$. Then $f(\bx)$ is the expected influence spread for seed set $\bx$\footnote{$f(\bx)=\sigma(S)$ if $x_j=1$ whenever $j\in S$, and 0 otherwise.}. We stress that $f(\cdot)$ is highly dependent on $\mathcal G$ as well as the diffusion process, but suppress this dependence in the notation for convenience. In practice, we cannot observe $f(\bx)$ directly; instead we obtain a noisy realization, $y$, via MC simulations, i.e.,
$$
    y = f(\bx) + \epsilon,
$$
where $y$ is the observed influence spread, $f(\bx)$ is the objective function and $\epsilon\sim\mathsf{Normal}(0,\sigma^2)$ is the noise term.

\subsection{Initial evaluations}
We initially obtain realizations of the spreading process for $N_0$ seed sets. We remark that nodes with high degrees on the temporally aggregated network, $\tilde G=\cup_t G_t$, are good candidates for the optimal seed set. Therefore, instead of randomly sampling $\bx_i$ from $\mathcal X$, we sample nodes {\it proportionally to their degree}. Mathematically, this means sampling $k$ times without replacement from a distribution with probability mass function $\mathsf{P}(Z=j)=d_j/\sum_k d_k$ where $d_j$ is the degree of node $j$ on $\tilde G$ for $Z\in\{1,\dots,n\}$. This sampling scheme ensures that the model better fits the objective function near the expected optimal seed set (see Supplemental Materials for empirical evidence). With $\bx_1,\dots,\bx_{N_0}$ sampled, we define ${\bf X}=(\bx_1^T,\dots,\bx_{N_0}^T)^T\in\{0,1\}^{N_0\times n}$ and obtain $\boldsymbol{y}=(y_1,\dots,y_{N_0})^T$ via MC simulations. Note that if identical seed sets are sampled, we discard one and re-sample to ensure $N_0$ unique initial seed sets. In the Supplemental Materials, we also discuss an alternative initial sampling scheme.

\subsection{Statistical model}
Next, we need an adequate surrogate model for the objective function. The most popular approach uses {\it Gaussian Process (GP)} regression. As the influence spreads are observed with noise, we assume the following GP regression model:
\begin{equation}\label{eq:gp}
    \boldsymbol{y}
    \sim\mathsf{Normal}(\mu({\bf X}), \sigma^2\gamma{\bf K} +\sigma^2 {\bf I}_{N_0} ),
\end{equation}
where $\mu(\cdot)$ is the mean function and ${\bf K}$ is the correlation matrix defined by a kernel function $\kappa(\cdot,\cdot)$ such that $K_{ij} = \kappa(\bx_i,\bx_j)$. Moreover, $\gamma>0$ is a kernel hyper-parameter, and $\boldsymbol{y}$ and ${\bf X}$ are defined in the previous sub-section. The two key components of the GP regression model are the mean function $\mu(\cdot)$ and kernel function $\kappa(\cdot,\cdot)$, so we consider each in turn.

\subsubsection{Mean function}\label{sec:mean}
First, we must specify the mean function $\mu(\cdot)$ of our GP regression model. As is typical, we consider an intercept-only model, i.e.,
$$
    \mu(\bx)
    =\beta_0{\bf 1}_{N_0}
$$
for all $\bx$, where $\beta_0$ is the intercept and ${\bf 1}_n$ is a vector of ones of length $n$. In Section \ref{sec:uq}, we introduce another mean function to allow for better uncertainty quantification of the optimal seed set.

\subsubsection{Kernel function}

To complete our GP regression model, we must also specify the correlation matrix, ${\bf K}$, defined by the kernel function, $\kappa(\cdot,\cdot)$. If $\bx_i,\bx_j\in\mathbb R^p$, then it is straightforward to compute their covariance using, e.g., a Gaussian or Matern kernel. In our setting, however, $\bx_i,\bx_j\in\{0,1\}^n$ and correspond to nodes in a graph, so it is not immediately obvious how to choose a kernel. We discuss several possible choices.

\paragraph{COMBO}
Recently, $\mathsf{COMBO}$ was proposed as an efficient and scalable method for combinatorial Bayesian optimization \citep{oh2019combo}. This paper proposes a novel kernel function for combinatorial search spaces based on the idea of a {\it combinatorial graph}, where each vertex\footnote{To attempt to avoid confusion between the combinatorial graph and the original graph where the information diffusion occurs, we will refer to the former as being made up of vertices, and the latter as possessing nodes.} corresponds to a possible input to the objective function, and there is an edge between vertices if the inputs differ by only a single variable. The authors construct such a graph using the graph Cartesian product and derive a kernel based on the Graph Fourier Transform.

In the context of IM, each vertex in the combinatorial graph would correspond to a possible seed set and there would be an edge between vertices if the corresponding seed sets only differed by one node. While we can construct such a graph, the cardinality constraint imposed on the seed sets means that we cannot construct the graph using the graph Cartesian product. Therefore, it is not straightforward to apply the ideas of $\mathsf{COMBO}$ to the IM problem. Please see the Supplemental Materials for further discussion of why the method is not applicable. We consider extending $\mathsf{COMBO}$ to constrained input spaces an interesting avenue of future work.

\paragraph{Hamming Distance}
In \cite{oh2019combo}, the authors show that the shortest distance between nodes on the combinatorial graph is equivalent to the Hamming distance. While we showed that $\mathsf{COMBO}$ cannot be used in our problem, we can borrow these ideas to construct a kernel function based on the Hamming distance. Specifically, let $\boldsymbol{x}_i$ and $\boldsymbol{x}_j$ be two seed sets. Then the covariance matrix is defined by
\begin{equation}\label{eq:sigmaH}
    K_{ij}
    =\kappa(\bx_i,\bx_j)
    =\begin{cases}
        1-\frac{1}{2k}d_H(\boldsymbol{x}_i,\boldsymbol{x}_j)&i\neq j\\
        1 + \delta & i=j
    \end{cases}
\end{equation}
where $\delta>0$ is some small constant to ensure positive-definiteness of our covariance matrix. In all experiments, we set $\delta=0.01$. Additionally, $d_H(\boldsymbol{u},\boldsymbol{v})$ is the Hamming distance, which counts the number of entries of between vectors $\boldsymbol{u}$ and $\boldsymbol{v}$ which are not equal, i.e.,
$$
    d_H(\boldsymbol{u},\boldsymbol{v})
    =\sum_{i=1}^n \mathbb I(u_i\neq v_i)
$$
where $\mathbb I(\cdot)$ is the indicator function. Because we are restricted to $k$ nodes in the seed set, if $k\leq n/2$, then $0\leq d_H\bx_i,\bx_j)\leq 2k$, which is why we divide by $2k$ instead of $n$. This kernel is in a similar spirit to those of $\mathsf{CASMOPOLITAN}$ \citep{wan2021think}, $\mathsf{Bounce}$ \citep{papenmeier2023bounce} and $\mathsf{CoCaBO}$ \citep{ru2020bayesian}. Please see the Supplemental Materials for a proof that this kernel is positive semi-definite.

\paragraph{Jaccard Coefficient}
While the Hamming distance is a sensible choice for a kernel metric, a potential drawback is that it does not account for the structure of the graph. Instead, it simply enumerates the number of nodes shared between seed sets. But since the graph structure, node neighbors, etc.~affect the influence spread, our kernel function should seemingly also account for these properties.


Recall that for the IM task, we initially influence a small set of nodes which then propagate information to the rest of the network. Clearly, these initially selected nodes (as well as their neighboring nodes) are very important for the final influence spread \citep{erkol2020influence}. Indeed, if the nodes corresponding to seed sets $\bx_i$ and $\bx_j$ have many neighbors in common, then we would expect their influence spreads also to be similar. 

Inspired by this observation, as well as \cite{garnett2010bayesian}, who studies the BO task on combinatorial search-spaces, we propose the following kernel function. Let $\{v_{i_1},\dots,v_{i_k}\}$ correspond to seed set $\bx_i$. First, we find all neighbors of $\{v_{i_1},\dots,v_{i_k}\}$ at time $t=1$, and call this neighboring set $\mathcal S_i$. In other words, if node $v_\ell$ has an incoming edge from some node in $\{v_{i_1},\dots,v_{i_k}\}$ at time $t=1$, then $v_\ell\in \mathcal S_i$. Of course, $\{v_{i_1},\dots,v_{i_k}\}\in\mathcal S_i$ as well.  We also compute the set of neighboring nodes set, $\mathcal S_j$, for $\bx_j$. Again, if $\mathcal S_i$ and $\mathcal S_j$ have many common elements, then we want $\kappa(\bx_i,\bx_j)$ to be large. Therefore, we can compute the Jaccard coefficient between $\mathcal S_i$ and $\mathcal S_j$ and use this as our kernel function. The Jaccard coefficient (JC) \citep{jaccard1901etude, jaccard1912distribution} is a simple metric to quantify the similarity between two finite sets. If $A$ and $B$ are two such sets, then the JC is the number of common elements between the two sets divided by the total number of elements in each set, i.e.,
$$
    JC(A,B)
    =\frac{|A\cap B|}{|A\cup B|}.
$$
If $A$ and $B$ are the same set, then $JC(A,B)=1$ and if $A$ and $B$ have no common elements, then $JC(A,B)=0$ such that $0\leq JC(A,B)\leq 1$. Additionally, it is clear that $JC(A,B)=JC(B,A)$.

With kernel function in hand, we can easily construct our covariance matrix:
\begin{equation}\label{eq:sigmaJ}
    K_{ij}
    =\kappa(\bx_i,\bx_j)
    =\begin{cases}
        JC(\mathcal S_i,\mathcal S_j)&i\neq j\\
        1 + \delta & i=j
    \end{cases}
\end{equation}
where again, $\delta>0$ is some small constant ($\delta=0.01$). This definition satisfies the {\it desideratum} that points close in the input space are more strongly correlated where we have defined two seed sets to be ``close'' if they have many neighbors in common at time $t=1$.

The use of the JC in such context enjoys some precedence. First, the JC has been used for link prediction in graphs, most notably in \citep{liben2003link}. Additionally, \cite{gosnell2024gaussian} use the Tanimoto index, another name for the JC, to construct a GP kernel with drug discovery applications. More generally, the JC has received significant attention in chemoinformatics \citep{rogers1960computer, gower1971general, todeschini2012similarity, moss2020gaussian, gosnell2024gaussian}. Finally, the JC (Tanimoto index) has been shown to yield a positive semi-definite correlation matrix, making it a valid choice for our kernel function \citep{gower1971general, bouchard2013proof}. 


\subsection{Acquisition function}
With our GP model fully defined, we can move on to the final step in the BO algorithm: the acquisition function. This function determines which seed set to evaluate the objective function at next. We want to probe areas of the search space with a high likelihood of being near the global maximum (exploitation), while also considering seed sets which have not been explored yet to prevent getting stuck in a local optimum (exploration). Moreover, this function will also need to account for the fact that we observe the influence spreads with noise.

We prefer the Expected Improvement acquisition function \citep{movckus1975bayesian, jones1998efficient, frazier2018tutorial} as it is one of the most popular acquisition functions and does not require any hyper-parameter tuning. The idea is to evaluate the objective function at the value which leads to the largest expected increase compared to the current maximum. Specifically, assume that we are in $b$-th iteration of the BO loop. Let $\bx\in\{0,1\}^n$ and $\tilde\mu^{(b)}(\bx)$ and ${\tilde \sigma}^{(b)}(\bx)$ be the conditional mean and standard deviation, respectively, of our GP model given the current data, where we explicitly emphasize the dependence on the iteration with the super-script in the notation. By properties of the multivariate normal distribution, we have
$$
    \tilde\mu^{(b)}(\bx)
    =\beta_0^{(b)} + \kappa^*(\bx, {\bf X})^T(\gamma{\bf K}+{\bf I}_{N_0+b})^{-1}(\boldsymbol{y}-\beta_0^{(b)}{\bf 1}_{N_0+b})
$$
and
$$
    {\tilde\sigma}^{(b)}(\bx)
    ={\sigma}^{(b)}\{(\gamma+1)-\kappa^*(\bx, {\bf X})^T(\gamma{\bf K}+{\bf I}_{N_0+b})^{-1}\kappa^*(\bx, {\bf X})\}^{1/2}
$$
where
$$
    \kappa^*(\bx,{\bf X})
    =(\kappa(\bx, \bx_1),\dots,\kappa(\bx, \bx_{N_0+b}))^T.
$$
In practice, we replace $\beta_0^{(b)}$ and ${\sigma}^{(b)}$ with their corresponding posterior medians. If our observations were noiseless, we would let $\Delta^{(b)}(\bx)=\tilde\mu^{(b)}(\bx)-{f^{(b)}}^*$ where ${f^{(b)}}^*=\max_{i\in\{1,\dots,N_0+b}\{y_i\}$, i.e., the maximum influence spread of the seed sets we have already evaluated. Then
\begin{equation}\label{eq:ei}
    EI^{(b)}(\bx)
    =\max\{\Delta^{(b)}(\bx),0\} + \tilde\sigma^{(b)}(\bx)\phi\left(\frac{\Delta^{(b)}(\bx)}{\tilde \sigma^{(b)}(\bx)}\right) - |\Delta^{(b)}(\bx)| \Phi\left(\frac{\Delta^{(b)}(\bx)}{\tilde\sigma^{(b)}(\bx)}\right)
\end{equation}
where $\phi(\cdot)$ and $\Phi(\cdot)$ are the probability density function and distribution function, respectively, of the standard normal \citep{frazier2018tutorial}. Recall, however, that the influence spread, $y_i$, is observed with Gaussian noise $\varepsilon_i$, making ${f^{(b)}}^*$ a (potentially) poor and non-robust estimate of the global maximum, particularly if it was observed with large noise. Following \cite{vazquez2008global} and \cite{picheny2013benchmark} , we replace ${f^{(b)}}^*$ with a ``plug-in'' estimate of the global maximum; namely, the maximum of the conditional mean function, $\tilde\mu^{(b)}(\cdot)$, evaluated at the previously sampled points, i.e.,
$$
    {f^{(b)}}^*
    =\max_{i\in\{1,\dots,N_0+b\}}\tilde\mu^{(b)}(\bx_i).
$$

We are still not quite finished as the acquisition function in \eqref{eq:ei} implies that we observe the next observation deterministically. To account for the noise in the ensuing observations, we included a multiplicative term as in \cite{huang2006sequential, huang2006global}, which penalizes sampling a new seed set where the conditional standard deviation, ${\tilde\sigma}^{(b)}(\cdot)$ is small. This is a sensible choice, as a small variance implies we already know the influence spread of this seed set with fairly high certainty, encouraging exploring elsewhere in the input space. The final (augmented) expected improvement (AEI) acquisition function becomes
\begin{equation}
    AEI^{(b)}(\bx) = EI^{(b)}(\bx)\times\left(1 - \frac{{\sigma^2}^{(b)}}{\sqrt{\{\tilde\sigma^{(b)}(\bx)\}^2+{\sigma^2}^{(b)}}}\right),
\end{equation}
and we choose the next seed set to evaluate, $\tilde{\bx}$ as the seed set with largest expected improvement, i.e.,
$$
    \tilde{\bx}={\arg\max}_{\bx\in\{0,1\}^n} AEI^{(b)}(\bx).
$$

The astute reader will notice that we have simply traded one combinatorial optimization problem (maximizing $f(\cdot)$ for maximizing $AEI(\cdot)$). $AEI(\cdot)$, however, is much cheaper to evaluate than $f(\cdot)$ so finding its optimum will be significantly faster. Indeed, we propose a simple greedy algorithm to find the maximum of $AEI(\cdot)$ with details in the Supplemental Materials. The main idea of this algorithm is that nodes in the current seed set are swapped with other nodes. If the switch improves the solution (larger $AEI(\cdot)$), then its kept, otherwise the previous seed set is used. In this way, the algorithm makes the locally optimal decision. The algorithm continues until all nodes have been looped through and no changes are made. Please see the Supplemental Materials for a connection between the proposed greedy algorithm and well-known Trust region (TR) framework \citep[e.g.,][]{eriksson2019scalable, wan2021think, papenmeier2023bounce}

After solving for $\tilde{\bx}=\arg\max AEI^{(b)}(\bx)$, we append it and its evaluated influence spread to ${\bf X}$ and $\boldsymbol{y}$, respectively, update the covariance matrix ${\bf K}$, and re-fit the GP regression model to update the posterior distribution of $\beta_0$ and $\sigma^2$. This step is repeated $B$ times where $B$ is the number of iterations in the BO loop. Then the optimal seed set is
$$
    \bx^*=\arg\max_{\bx\in\{\bx_1,\dots,\bx_{N_0+B}\}}\tilde\mu^{(B)}(\bx),
$$
where this maximum is taken over the $N_0+B$ sampled points and depends on the final posterior distribution of $\beta_0$ and $\sigma^2$.

The steps for the entire algorithm are outlined in Algorithm \ref{alg:app}. For a given network $\mathcal G$, diffusion mechanism, seed size $k$, and budget $N_0,B$, the algorithm outputs the optimal seed set to maximize the influence spread on $\mathcal G$. We call the method $\mathsf{BOPIM}$ for Bayesian OPtimization for Influence Maximization on temporal networks.

\begin{algorithm}
\SetAlgoLined
\KwResult{Optimal seed set $S$}
 {\bf Input: } Objective function $f(\cdot)$; temporal network $\mathcal G$; seed set size $k$; budget $B$; number of initial samples $N_0$; \;

\For{$N_0$ times}{
    Sample $\bx_i$ proportional to node's aggregate degree\;
    
    Evaluate $y_i=f(\bx_i)$\;
}

Construct ${\bf K}$ using \eqref{eq:sigmaH} or \eqref{eq:sigmaJ} \;

Compute posterior of $\beta_0$ and $\sigma^2$\;

\For{ $B$ times}{

    Find $\tilde{\bx}=\arg\max_{\bx, \sum_j x_j=k} AEI(\bx)$ \;

    Compute $y=f(\tilde{\bx})$\;

    Append observation and re-compute posterior of $\beta_0$ and $\sigma^2$\;

    Update ${\bf K}$ \;

}

 \Return{$\bx^*=\arg\max_{\bx\in\{\bx_1,\dots,\bx_{N_0+B}\}}\tilde\mu^{(B)}(\bx)$}
 
 \caption{Bayesian optimization for influence maximization: $\mathsf{BOPIM}$}
 \label{alg:app}
\end{algorithm}

\subsection{Discussion}
Finding the seed set for IM on temporal networks is a cardinality-constrained, combinatorial optimization problem with non-Euclidean inputs. We primarily address these challenges with the kernel function and acquisition function. Both of these components of our BO algorithm are specifically tailored to handle the unique challenges posed by IM. 

$\mathsf{BOPIM}$ is much faster than a standard greedy approach since we only need to obtain $N_0+B$ realizations of the costly objective function, as opposed to $O(nk)$ times for a greedy algorithm. Thus, the number of evaluations of the objective function for $\mathsf{BOPIM}$ is independent of $k$. Moreover, it is trivial to adapt the algorithm to other diffusion mechanisms. In practice, the diffusion model should be chosen with great care based on the application at hand. But since the diffusion model only appears in the evaluation of the objective function, and the objective function is treated as a ``black-box'', changing the diffusion model has no impact on the implementation of the method. This differs from, e.g., Dynamic Degree \citep{murata2018extended}, which only works for the SI model, or \cite{erkol2020influence}, which is based on the SIR model.

One of the main advantages of the BO framework is that it easily allows for quantifying uncertainty. We discuss this in more detail in Section \ref{sec:uq}, but briefly anticipate that discussion. By evaluating a given seed set using the posterior distribution of $\beta_0$ and $\sigma^2$, we obtain a complete posterior predictive distribution for the total influence spread. This is in contrast with node-ranking heuristics, which do not yield an estimated influence spread, and probabilistic approximations, which yield a point estimate but not a measure of uncertainty of the total spread. Moreover, if a researcher wants to estimate the expected influence spread for another seed set $S'\neq S$, then the entire calculation needs to be re-run for previous methods. After $\mathsf{BOPIM}$ has been fit once, the influence spread for any seed set can be estimated immediately and the uncertainty in the optimal seed set can also easily be quantified.

\subsection{Hyper-parameter selection}
Finally, there are several hyper-parameters that need to be chosen for $\mathsf{BOPIM}$. Choosing $\gamma$, the kernel hyper-parameter, is perhaps the most tricky. We note that $\gamma$ corresponds to the contribution of ${\bf K}$ to the overall variance of the GP regression model. If $\gamma$ is small, then most of the variance is simply coming from the random noise term, $\epsilon_i$, but if $\gamma$ is large, then the variance in the objective function evaluations is largely driven by our kernel function. There are many different ways that we could choose $\gamma$. If we want a fully Bayesian approach, then we could endow $\gamma$ with a prior and update its posterior distribution during the MCMC routine. In practice, however, we found that this did not lead to good performance and required adding a Metropolis-Hastings step to an otherwise entirely Gibbs sampler. Another choice is to fix $\gamma$ at its maximum likelihood estimate (MLE). In our experiments, this choice proved to perform better than the fully Bayesian approach, but the MLE estimate of $\gamma$ was consistently close to 1. Indeed, instead of finding the MLE for $\gamma$ after each new observation, simply setting $\gamma=1$ yielded the best performance. This choice corresponds to equal contribution from ${\bf K}$ and the random error term.

\begin{figure}
    \centering
    \includegraphics{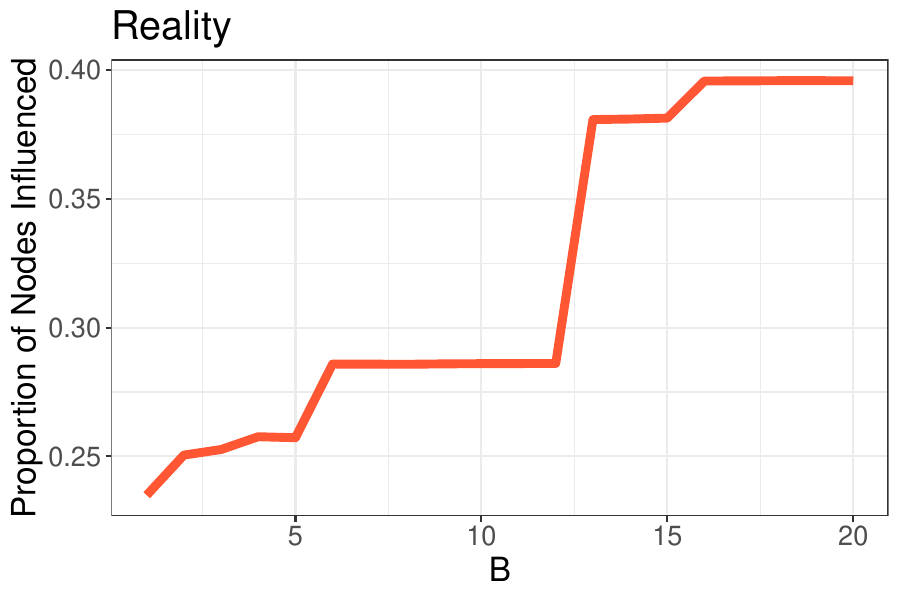}
    \caption{Progress curve of $\mathsf{BOPIM}$ algorithm for Reality network.}
    \label{fig:progress}
\end{figure}

There are several other parameters that must be chosen in our algorithm. For the number of initial samples $N_0$ and BO loop iterations $B$, if these values are large, then the surrogate function will fit the true function well, but at a high computational cost. We found that setting $N_0=5$ and $B=20$ led to a good balance of fit and speed. To show that the BO iterations improve the solution quality, we report a progress curve in Figure \ref{fig:progress} for a single run of the $\mathsf{BOPIM}$ algorithm. Please see Section \ref{sec:exp} for a description of the network. As the number of BO iterations increases, the total influence spread also increases. Indeed, for this particular example, the solution quality (proportion of nodes influenced) increases by over 16\% during the BO loop, which corresponds to a relative increase of almost 70\%. 

We use a standard Gibbs sampler to estimate the posterior distribution of $\beta_0$ and $\sigma^2$, meaning the number of MCMC samples must also be chosen. More MCMC samples leads to better estimates of the parameters, but also longer run-times. Since we only use the median of the posterior distribution, we can keep the number of MCMC samples relatively small to increase speed. We recommend 2500 iterations with the first 500 used for burn-in. Lastly, for the number of MC simulations for the objective function realization, it is important that this output is accurate so we recommend 1,000 iterations. Unless otherwise noted, we use these settings for all experiments in Section \ref{sec:exp}.

\section{Experiments}\label{sec:exp}
We now study the performance of the proposed method with numerical experiments. All timed experiments were performed in Python on a 2024 M4 Mac Mini with 16 GB of memory and 9 cores in use. The code for Algorithm \ref{alg:app} is available on GitHub: \url{https://github.com/eyanchenko/BOPIM}.

\subsection{Data}
The following experiments are conducted on four real-world networks. Reality ($n=64$, $m=722$) \citep{eagle2006reality}, Hospital ($n=75$, $m=1139$) \citep{vanhems2013estimating}, Bluetooth ($n=136$, $m=5949$) \citep{crawdad} and Conference 2 ($n=199$, $m=1550$) \citep{chaintreau2007impact} are all proximity networks where an edge exists if two people are close to each other. 
In order to discretize the continuous appearing and disappearing of edges, each network is aggregated into $T$ temporal snapshots of equal duration. We note that these networks are relatively small due to the slow speed of the greedy algorithm. Therefore, in the Supplemental Materials, we also conduct an experiment using the proposed method on the DNC email network \citep{nr} with $n=1891$ nodes to demonstrate its scalability. Additionally in the Supplemental Materials, we also quantify the fit of the surrogate model to the true objective function.

\subsection{Settings}

We simulate the influence diffusion process and select seed sets for IM using $\mathsf{BOPIM}$. The main metrics of comparison are the proportion of nodes influenced at the conclusion of the spreading process, as well as the total algorithmic run time. We consider the proposed algorithm with both Hamming and Jaccard kernel, denoted $\mathsf{BOPIM-H}$ and $\mathsf{BOPIM-J}$, respectively, with $N_0=5$ and $B=20$. A greedy algorithm is also used, considered the ``gold standard'' in IM. \cite{erkol2022effective} proved that the SI model of diffusion is monotone and sub-modular so we use the CELF \citep{leskovec2007cost} step to increase the speed of the greedy algorithm. We also compare with random sampling seed nodes (Random), randomly sampling proportional to a nodes' degree (Random Degree) and Dynamic Degree \citep{murata2018extended}.

All experiments are conducted under the {\it ex post} assumption where the future evolution of the network is known when selecting seed nodes as in \cite{murata2018extended, osawa2015selecting, aggarwal2012influential, erkol2020influence}. While this assumption may be unrealistic in some cases, the main goal of this work is to propose a BO scheme for IM rather than tackling the {\it ex ante} problem \citep{yanchenko2023link}. We leave this as an important avenue of future work. Finally, to compare the different methods, we vary the number of seed nodes $(k)$, granularity of snapshots $(T)$, and infection parameter $(\lambda)$. Results are averaged over 25 Monte Carlo (MC) replications (save Greedy and Dynamic Degree), and the influence spread results are reported with error bars.

\subsection{Results}

\begin{figure*}[!t]
    \centering
    \includegraphics[width=\textwidth]{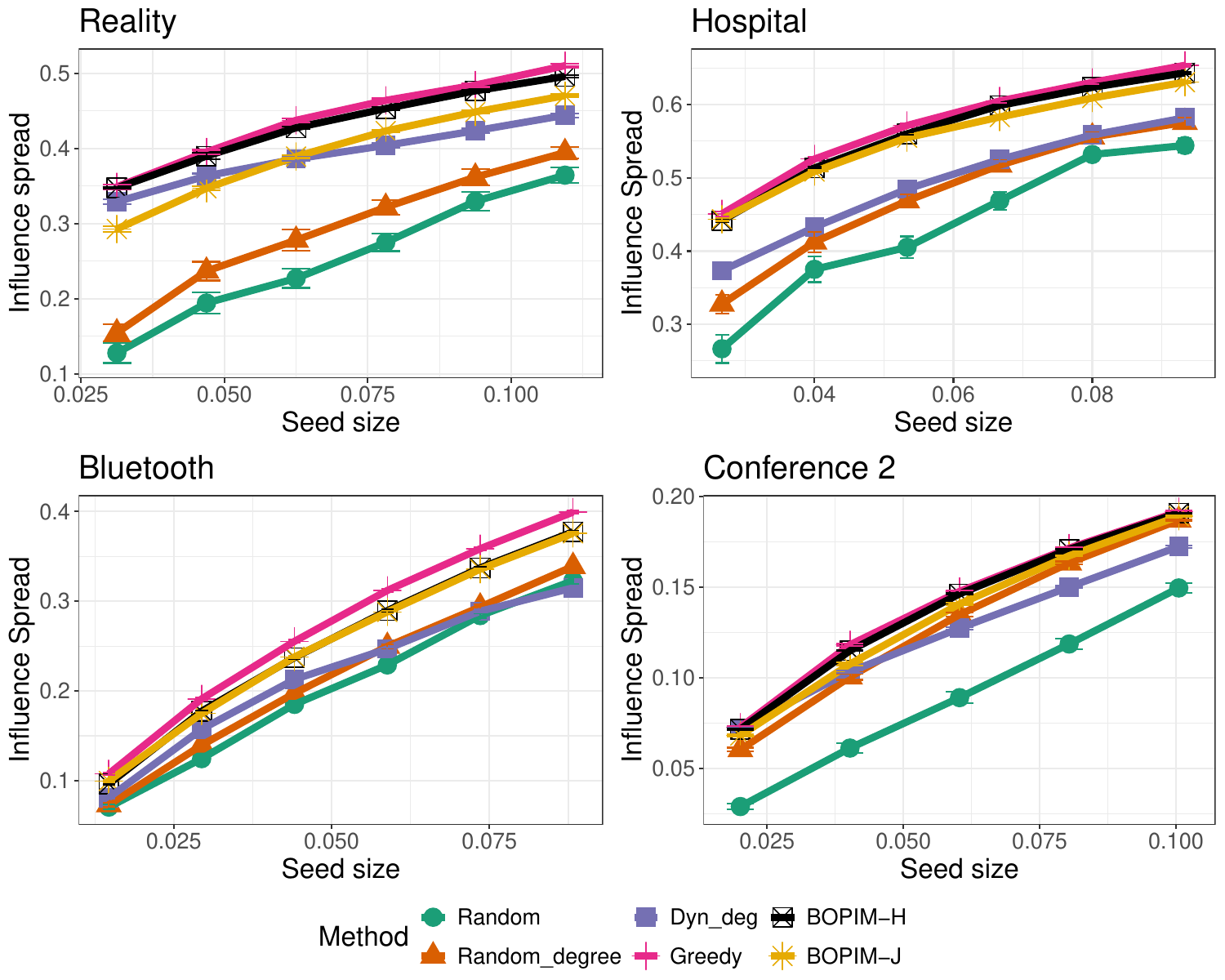}
    \caption{Influence spread results for increasing seed size $(k)$.}
    \label{fig:k}
\end{figure*}

\begin{figure*}[!t]
    \centering
    \includegraphics[width=\textwidth]{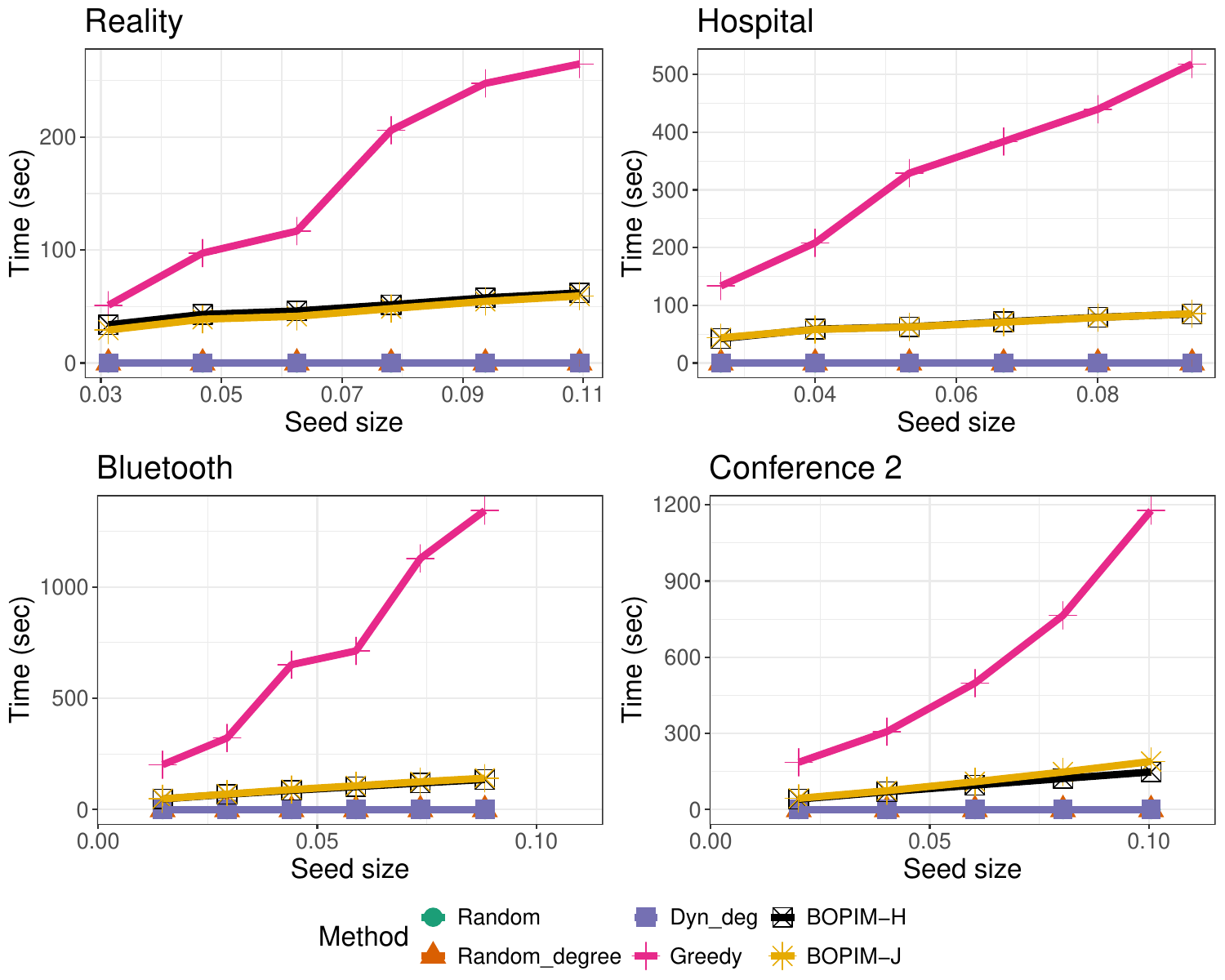}
    \caption{Computation time results for increasing seed size $(k)$.}
    \label{fig:time}
\end{figure*}

First, we fix $T=10$ and $\lambda=0.05, 0.05, 0.01, 0.01$ for the Reality, Hospital, Bluetooth and Conference 2 datasets, respectively, while varying $k$. In practice, $\lambda$ should be carefully chosen alongside domain experts, based on the particular scenario, but for illustrative purposes, we have chosen these values. For this setting, we record both the influence spread and computation time to select the seed set, with the results in Figures \ref{fig:k} and \ref{fig:time}. 

For each data set, the proposed BO method yields comparable influence spread to the greedy algorithm, especially with the Hamming kernel. For Reality, Hospital and Conference 2, in particular, the spreads are almost indistinguishable between $\mathsf{BOPIM-H}$ and greedy. Both $\mathsf{BOPIM}$ methods yield similar influence spreads, except for the Reality network, and also exceed those of Dynamic Degree in all settings. While Random Degree performs similarly to Dynamic Degree for some settings, Random always performs the worst.  In terms of computing time, the proposed method has a large advantage over the greedy algorithm. In each case, the computational time greatly increases with $k$ for the greedy algorithm, whereas that of the proposed method increases more slowly. For large $k$, the proposed method is as much as ten times faster than the greedy algorithm (Bluetooth network).

\begin{figure*}[!t]
    \centering
    \includegraphics[width=\textwidth]{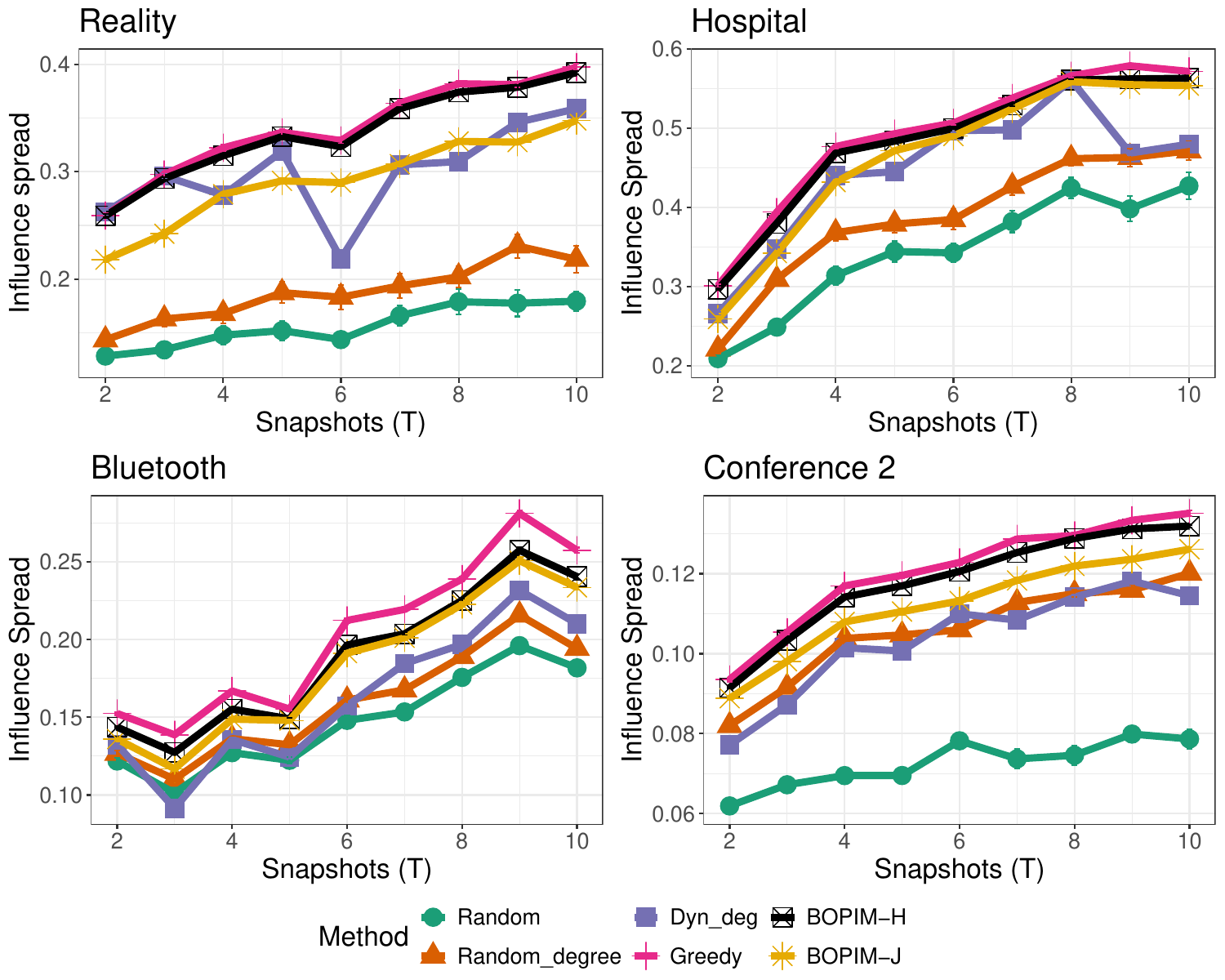}
    \caption{Influence spread results for increasing number of snapshots $(T)$.}
    \label{fig:snapshots}
\end{figure*}

Keeping the same values of $\lambda$ as before, we now fix $k$ such that roughly 5\% of the total number of nodes are initially activated. We vary the number of aggregations of the temporal network, where small and large $T$ correspond to granular and fine aggregations, respectively. While the ``real-time'' that the network diffuses remains the same, we can vary the number of aggregations to see how this affects the influence spread. The results are in Figure \ref{fig:snapshots}. 

The trends are similar to that of varying $k$. Across networks and $T$, the proposed method using Hamming kernel yields comparable influence spread to that of the greedy algorithm, again performing especially well in the Reality, Hospital and Conference 2 networks. We also see that the Hamming kernel noticeably outperforms the Jaccard kernel in the Reality, Hospital and Conference 2 networks. Dynamic Degree and Random Degree yield consistently smaller influence spreads than both $\mathsf{BOPIM-H}$ and Greedy.  In general, the total spread increases with $T$ as there are more rounds for influence to diffuse, but this trend need not be monotonic as different aggregations changes when certain nodes have edges. Indeed, in the Reality network, Dynamic Degree's results are quite sensitive to $T$ as compared to the proposed method.

\begin{figure*}[!t]
    \centering
    \includegraphics[width=\textwidth]{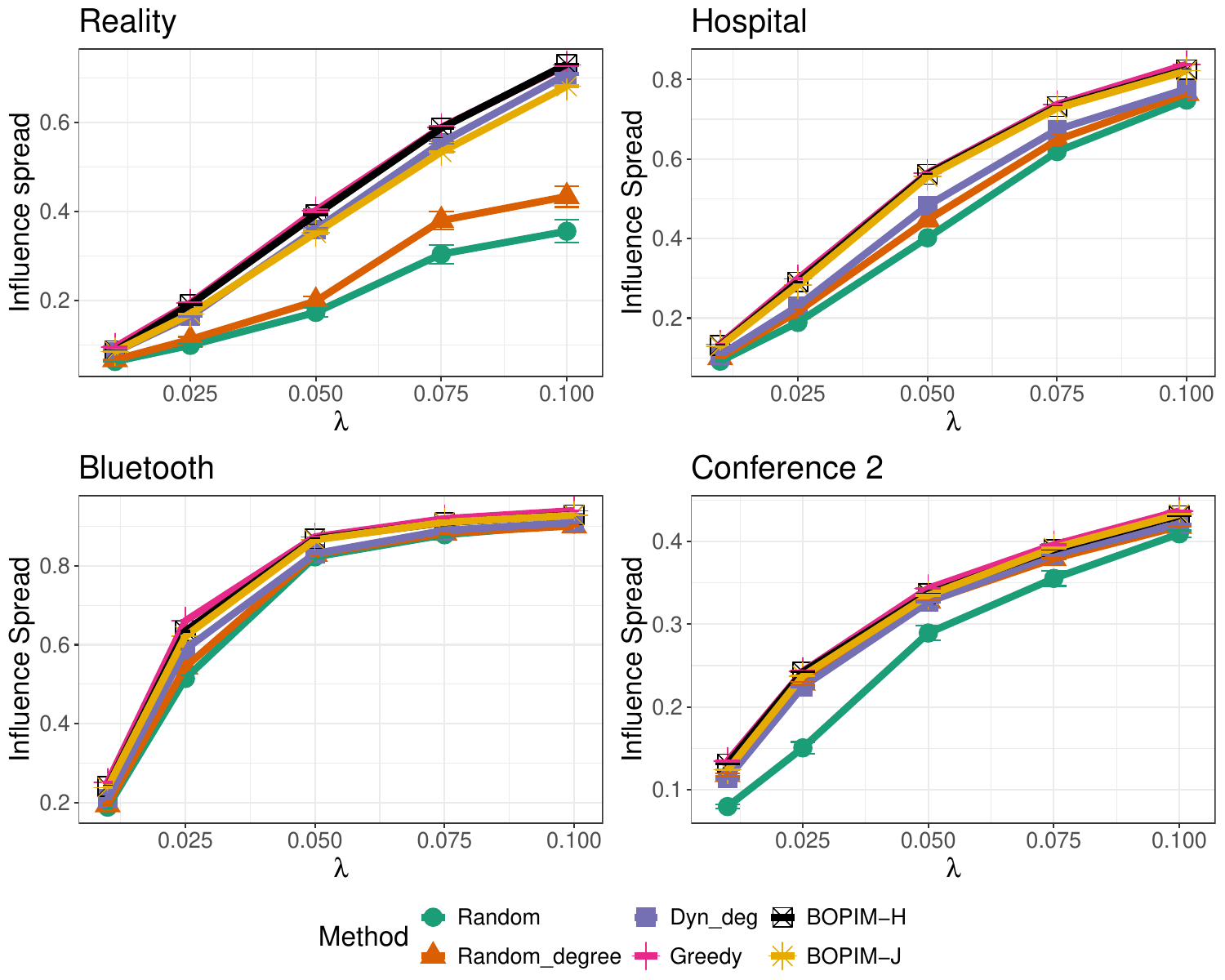}
    \caption{Influence spread results for increasing infection probability $(\lambda)$.}
    \label{fig:lambda}
\end{figure*}

Finally, we fix $k$ at about 5\% of $n$ and $T=10$ and vary the infection probability $\lambda$. The results are in Figure \ref{fig:lambda} and are similar for each network: as $\lambda$ increases, the total influence spread also increases. The distinction between different methods is least pronounced in this setting, but both versions of $\mathsf{BOPIM}$ have almost identical influence spreads to Greedy in all settings and networks.

\section{Uncertainty quantification}\label{sec:uq}

One of the major advantages of the BO framework is that it naturally provides uncertainty quantification (UQ) results. Indeed, $\mathsf{BOPIM}$ is one of the first such methods for the IM problem. UQ allows for a richer understanding of the outputted solution and can answer questions such as: Are there many seed sets which yield near optimal spread? Can the selection of seed nodes be viewed on a spectrum as opposed to binary inclusion/exclusion? We take a first step towards answering these questions with the proposed method in this section by discussing two ways that $\mathsf{BOPIM}$ can be leveraged for UQ.

\subsection{Posterior distribution of node contributions}

\subsubsection{Modified mean function}
Recall that the mean function in Section \ref{sec:mean} only included an intercept term. While this yielded seed sets with large influence spread, if we are interested in uncertainty quantification, it can be helpful to consider a more richly parameterized mean function. 

Specifically, let the mean function $\mu(\cdot)$ be parameterized by $\bbeta$. Motivated by \cite{baptista2018bayesian}, we assume that the surrogate model is linear in its parameters, i.e.,
\begin{equation}\label{eq:surr}
    \mu(\bx_i)
    =\sum_{j=1}^n x_{ij}\beta_j
    =\bx_i\bbeta
\end{equation}
for $\bbeta=(\beta_1,\dots,\beta_n)^T$. GP regression places a multivariate normal prior $\bbeta\sim\mathsf{Normal}({\bf 0}_n,\Sigma_\beta)$ where ${\bf 0}_n$ is the vector of zeros of length $n$ and $\Sigma_\beta$ is some covariance matrix. We also empirically standardize $\boldsymbol{y}$ to have mean zero to remove the need for an intercept.

Since $n$ is large for most real-world networks, the dimension of $\bbeta$ will also be large. Additionally, it is unlikely that each component of $\bbeta$ is significant, i.e., many nodes inclusion in the seed set does not lead to significant influence spread. Thus, we endow $\bbeta$ with a {\it sparsity-inducing shrinkage prior}. As in \cite{baptista2018bayesian}, we use the Horseshoe (HS) prior \citep{carvalho2009handling, carvalho2010horseshoe} which is designed to shrink insignificant coefficient estimates towards zero while still allowing for accurate estimation of the important coefficients. Specifically, the HS prior handles sparsity via a half-Cauchy prior distribution on the variance of the coefficients, i.e., 
\begin{align}\notag
    \beta_j|\lambda_j^2,\tau^2,\sigma^2&\sim \mathsf{Normal}(0, \eta_j^2\tau^2\sigma^2)\\
    \eta_j^2,\tau^2&\sim\mathsf{C}^+(0,1)\
\end{align}
where $j=1,\dots,n$ and $\mathsf{C}^+(0,1)$ is the standard half-Cauchy distribution. Thus, $\tau^2$ represents the global shrinkage and $\eta_j^2$ controls the local shrinkage of each coefficient. For the Gibbs sampler, we leverage an auxiliary variable formulation \citep{makalic2015simple}. Please see the Supplemental Materials for complete MCMC details. We also considered the R2D2 prior \citep{zhang2022bayesian, yanchenko2021r2d2} and Dirichlet-Laplace \citep{bhattacharya2015dirichlet}, but the results were similar, so we focus on the HS prior.

\subsubsection{Coefficient interpretation}
The coefficient $\beta_j$ represents the marginal importance of each node to the overall influence spread. This means that if the estimate of $\beta_j$ is positive and large, then including node $j$ in the seed set leads to a larger total influence spread. Conversely, small and negative coefficient estimates correspond to nodes whose inclusion in the seed set does not yield large influence spreads. We stress, however, that the coefficient estimates are {\it not} interpretable in the traditional sense. Typically, the coefficient $\beta_j$ for a binary explanatory variable $x_j\in\{0,1\}$ corresponds to the increase in the response when $x_j=1$ and all other variables are fixed. This interpretation is invalid in our context, however, due to the constraint $\sum_{j=1}^n x_j=k$. In other words, $\beta_j$ does not correspond to the increase in the influence spread if node $j$ is added to the seed set and all other nodes are held fixed, because then the number of nodes in the seed set would be greater than $k$. Instead, we interpret $\beta_j-\beta_i$ as the expected increase in the influence spread function if node $j$ replaced node $i$ in the seed set. Thus, the estimates yield a relative sense of the increase/decrease of the objective function if a node is seeded, and in this sense, are analogous to a node-ranking heuristic.

\subsubsection{Results}
We can leverage the posterior distribution of $\bbeta$ to quantify the uncertainty of a node's importance. To do this, we save the final posterior distribution of $\bbeta$ in $\mathsf{BOPIM}$ for the Reality network with $k=5$, $T=10$, $\lambda=0.05$ (6,000 MCMC samples, 1,000 burn-in) and Hamming kernel. In Figure \ref{fig:alpha_box}, we report the box-plot for the posterior distribution of $\beta_i$ for each node $i\in\{1,\dots,n\}$. The largest posterior distributions correspond with the ideal nodes for the seed set. From the figure, nodes 1, 2, and 3 have noticeably larger values of the posterior distribution than those of the other nodes. Thus, while $k=5$ nodes are chosen for the seed set, it may be that nodes 1, 2, and 3 are more influential than the other node included in the seed set.

\subsection{Proportion of iterations selected for seed set}
\begin{figure*}[!t]
    \centering
     \begin{subfigure}[b]{0.8\textwidth}
         \centering
         \includegraphics[width=\textwidth]{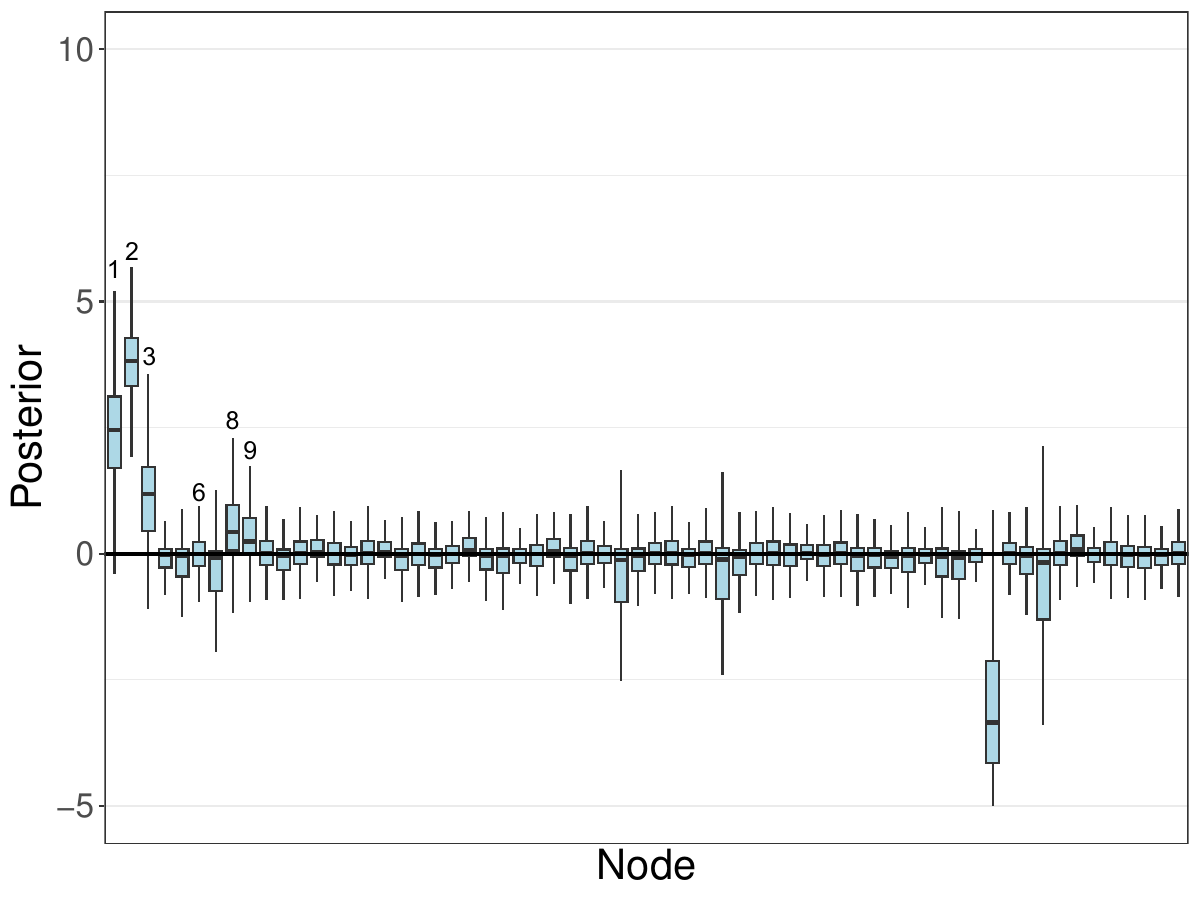}
         \caption{Posterior distribution box plots of $\bbeta$.}
         \label{fig:alpha_box}
     \end{subfigure}
     \\ 
     \begin{subfigure}[b]{0.8\textwidth}
         \centering
         \includegraphics[width=\textwidth]{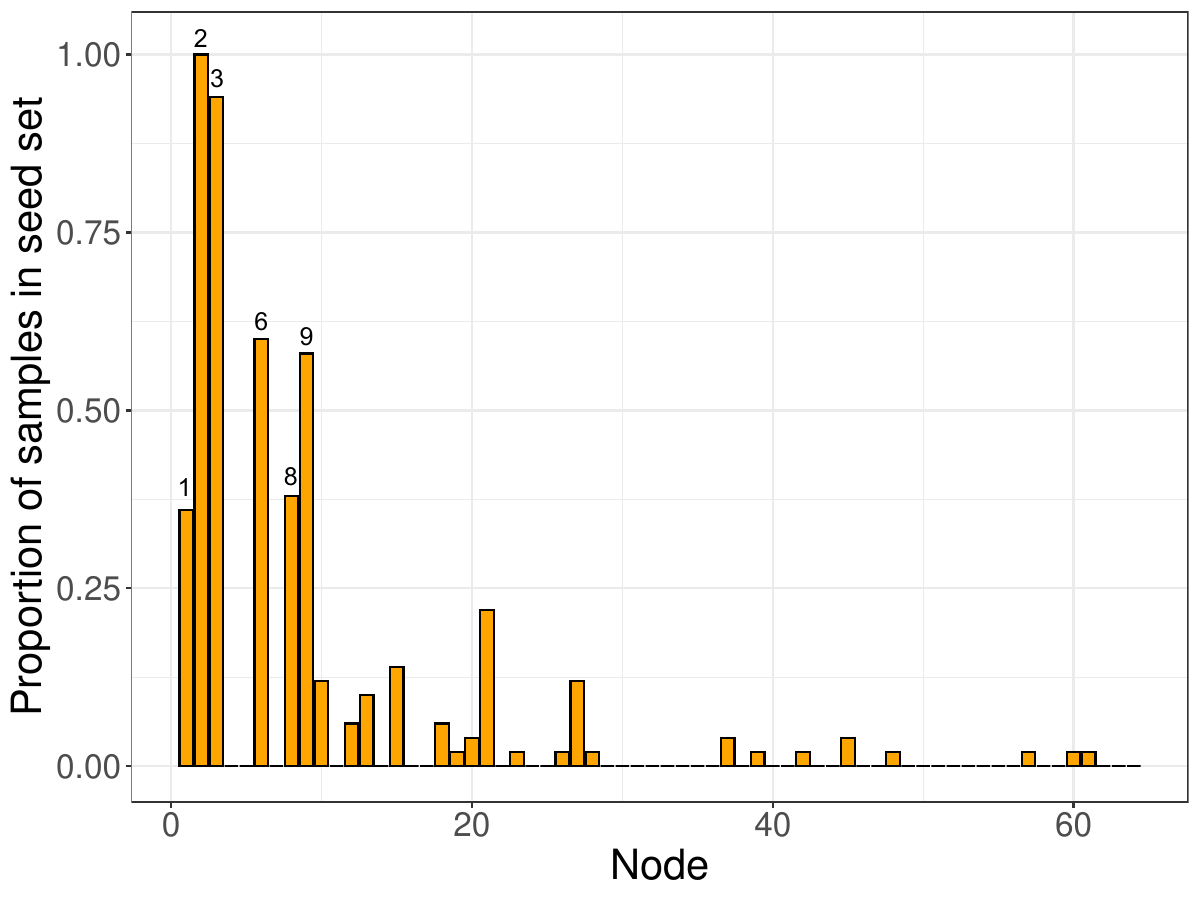}
         \caption{Proportion of MC samples selected for seed set.}
         \label{fig:alpha_prop}
     \end{subfigure}
        \caption{Posterior distribution summaries for the Reality network.}
        \label{fig:three graphs}
\end{figure*}

While $\mathsf{BOPIM}$ seeks the global optimum of our objective function, in practice, running the algorithm multiple times will give different optimal seed sets due to the randomness inherent in the algorithm. We can leverage this randomness as another way to quantify the uncertainty in the optimal seed sets. We now run the algorithm for 50 MC simulations (2,500 MCMC samples with 500 burn-in), and for each iteration, we find the optimal seed set. Then we average these results to find the proportion of times that each node is assigned to the optimal seed set. Note that this measure of uncertainty holds regardless of the mean function, so we revert back to the intercept-only model for the following experiment. These proportions are reported in Figure \ref{fig:alpha_prop}. For example, node 2 was selected in the seed set in 100\% of the iterations, thus making it the most influential node. While nodes 2 and 3 have noticeably larger proportions than all other nodes, the third through fifth largest (6, 8 and 9) have much smaller proportions than that of those two.

We can note some differences between the two methods of UQ. Since Figure \ref{fig:alpha_box} is based on a single iteration, the results are subject to more stochastic variability, whereas much of this effect is averaged out in Figure \ref{fig:alpha_prop}. For example, node 1 has the second largest posterior distribution, but is only the sixth most likely node to appear in the seed set when averaged over several iterations. On the other hand, the posterior mean of nodes 9 and 6 are close to zero, but they are the third and fourth most important nodes, respectively, when we re-run the algorithm many times. Regardless, these two figures together give us a good sense of the uncertainty in our optimal seed set and show that there are multiple sets of $k$ nodes which likely yield similar influence spreads.

\section{Conclusion}\label{sec:conc}

In this work, we apply a Bayesian Optimization framework to solve the influence maximization problem on temporal networks. The proposed $\mathsf{BOPIM}$ algorithm models the influence spread function via Gaussian Process regression. In particular, we propose two kernel functions, one based on the Hamming distance and the other on the Jaccard coefficient. We also adopted the Expected Improvement acquisition function, optimizing this with a greedy algorithm which accounts for the cardinality constraints. Finally, we demonstrated $\mathsf{BOPIM}$'s utility in uncertainty quantification (UQ).

In all experiments, the proposed method yields influence spreads comparable to that of seed sets from a greedy algorithm. The BO approach, however, is much as ten times faster than the greedy algorithm. Additionally, we consistently found that the Hamming kernel was equivalent to, or outperformed, the Jaccard kernel. This result is somewhat surprising as we expected the Jaccard kernel to perform better since it explicitly accounts for the graph structure, unlike the Hamming distance. Additionally, the UQ results gave rich insights into the importance of nodes in the seed set. In particular, we found evidence that there are many seed sets which yield comparably optimal influence spread, thus implying that the objective function is relatively ``flat.''

There are many interesting avenues to extend this work. First, we could study the performance of the proposed method under various forms of model mis-specification. Indeed, we take a first step in this direction in the Supplemental Materials where we consider two different cases. First, the algorithm is trained on one value of $k'$ but then the resulting model fit is used to find the optimal seed set for $k\neq k'$. In the second setting, we fit the algorithm on one value of $T'$, the number of temporal snapshots, and then use the resulting seed set to compute the influence spread on the network with $T\neq T'$ snapshots. In both cases, the $\mathsf{BOPIM}$ algorithm performs quite favorably. This not only implies a robustness of the method, but also can lead to computational advantages as the model may not need to be re-fit for every new parameter combination. These are important and encouraging findings, but there is certainly room for more study.

Another important area for future work is constructing the kernel function. Indeed, the surprising results of the Hamming distance's superior performance indicates that the role of the kernel function in this problem is not fully understood. One potential direction comes from Automatic Relevance Detection (ARD) \citep{wipf2007new, neal2012bayesian}. This approach enforces sparsity in the model via the covariance function and has already enjoyed success in BO papers \citep[e.g.,][]{papenmeier2023bounce}. Finally, considering the {\it ex ante} IM task is another important open-problem.


\section*{Supplemental Materials}

{\bf Supplemental methods and results}
\newline
Proof of positive semi-definiteness of Hamming kernel (Section 1); discussion of COMBO algorithm's applicability in IM (Section 2); greedy algorithm to optimize acquisition function (Section 3); relationship between greedy algorithm and Trust regions (Section 4); large network empirical results (Section 5); surrogate function validation results (Section 6); sampling scheme ablation study (Section 7); robustness experiments with $k$ and $T$ (Section 8); Horseshoe prior Gibbs sampler (Section 9)\\

\noindent
{\bf Python code and data}
\newline
Python code and data to replicate all results in Section \ref{sec:exp} and re-create Figures \ref{fig:k} - \ref{fig:lambda}. Code is also available on GitHub: \url{https://github.com/eyanchenko/BOPIM}.

\section*{Acknowledgments}
This work was partially conducted while the author was on a Japan Society for the Promotion of Science (JSPS) fellowship. This work was also supported by JSPS KAKENHI grant 24K22613. The author would like to thank Tsuyoshi Murata and Petter Holme for their mentorship, as well as the editors and anonymous reviewers whose comments greatly improved the quality of the manuscript. Finally, ChatGPT-4o was used in the Hamming kernel proof to notice the connection with Gram matrices.

\section*{Disclosures}
The author reports there are no competing interests to declare.

\bibliographystyle{apalike}
\bibliography{refs}

\clearpage

\begin{center}
    \Large Supplemental Materials
\end{center}

This Supplemental Materials contains the following sections:
\begin{enumerate}
    \item Proof of positive semi-definiteness of Hamming kernel
    \item Further discussion of the COMBO algorithm in relation to the IM problem
    \item The greedy algorithm to optimize the acquisition function
    \item Discussion on the relationship between the greedy algorithm and Trust regions
    \item Experiment comparing $\mathsf{BOPIM}$ with competing methods on a larger network
    \item Surrogate function validation results
    \item Ablation study on the sampling scheme for the initial observations in the BO algorithm
    \item Robustness experiments when varying the training/testing values of $k$ and $T$
    \item Horseshoe prior Gibbs sampler
\end{enumerate}

\clearpage

\section*{1. Proof of Hamming Kernel being positive semi-definite}

First, we assume that $k<\frac1{2}n$. Recall that $\kappa(\bu, \bv)=1-\frac1{2k}d_H(\bu,\bv)$ for $i\neq j$, and $1+\delta$ if $i=j$. We want to prove that this function is positive semi-definite (psd) for all $\bu,\bv\in\{0,1\}^n$ where $\sum_i u_i=\sum_i v_i=k$. 

\noindent
First, we can re-write the kernel as
$$
    \kappa(\bu,\bv)
    =\frac{2k-n}{2k} + \frac1{2k}\sum_{i=1}^n \mathbb I(u_i=v_i),
$$
and note that the value is always positive due to the cardinality constraint on the input vectors.
Then
$$
    \sum_{i=1}^n \mathbb I(u_i=v_i)
    =\sum_{i=1}^n \mathbb I(u_i=v_i=1) + \sum_{i=1}^n \mathbb I(u_i=v_i=0)
    =\bu^T\bv + ({\bf 1}_n-\bu)^T({\bf 1}_n-\bv)
$$
where ${\bf 1}_n$ is the vector of one's of length $n$. Thus, we can re-write the kernel as
\begin{align*}
    \kappa(\bu,\bv)
    &=\frac{2k-n}{2k} + \frac1{2k}\bu^T\bv + \frac1{2k}({\bf 1}_n-\bu)^T({\bf 1}_n-\bv)\\
    &=\frac{2k-n}{2k} + \frac1{2k}\bu^T\bv  + \frac{n}{2k} - \frac{k}{2k} - \frac{k}{2k} + \frac1{2k}\bu^T\bv \\
    &=\frac1k\bu^T\bv.
\end{align*}
Thus, the covariance matrix is a scaled Gram matrix, implying positive semi-definiteness. $\square$

\section*{2. COMBO algorithm}
\begin{figure}
    \centering
    \includegraphics[width=0.5\linewidth]{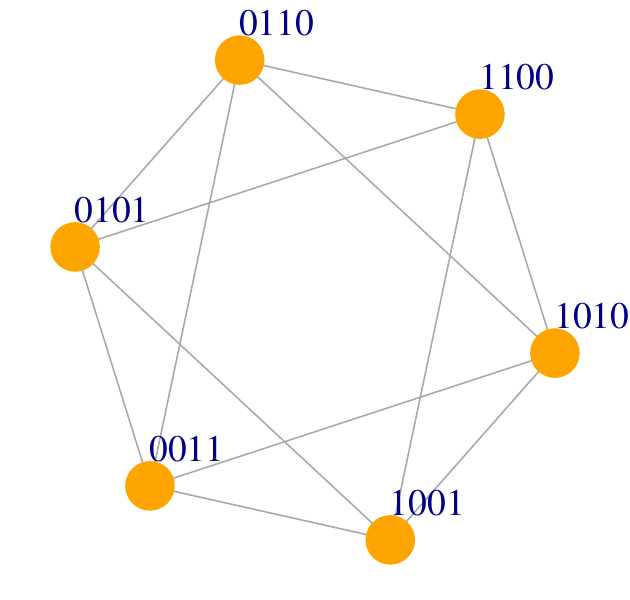}
    \caption{Combinatorial graph for possible seed sets with $n=4$ and $k=2$. Vertex label corresponds to the nodes assigned to the seed set where a 1 in position $i$ indicates that node $i$ is included in the seed set, and 0 otherwise. There is an edge between vertices which differ by only one node in the seed set.}
    \label{fig:combo1}
\end{figure}

In this section, we briefly discuss $\mathsf{COMBO}$ and further explain why it is inapplicable to our IM problem. The main idea behind $\mathsf{COMBO}$ is to construct a kernel function between combinatorial inputs based on the {\it combinatorial graph}.\footnote{To attempt to avoid confusion between the combinatorial graph and the original graph where the information diffusion occurs, we will refer to the former as being made up of vertices, and the latter as possessing nodes.} In this graph, each vertex represents a possible (combinatorial) input to the objective function of interest, and edges connect vertices which differ only by a single input. In our context, this means that each vertex of the graph represents a possible seed set, and there is an edge between vertices if they differ by a single node in their respective seeds sets. To fix ideas, we construct the combinatorial graph for a toy example where we have $n=4$ nodes and a seed set of size $k=2$. Then the possible seed sets are $S\in\{1100, 1010, 1001, 0110, 0101, 0011\}$ where the elements in $S$ have a 1 in position $i$ if node $i$ in included in the seed set, and 0 otherwise for $i=1,2,\dots,n=4$. For example, $1100$ means that nodes 1 and 2 are in the seed set. Then there is an edge between vertices if they share a seed node. The combinatorial graph can be seen Figure \ref{fig:combo1}.

\begin{figure}
        \centering
        \includegraphics[width=1\linewidth]{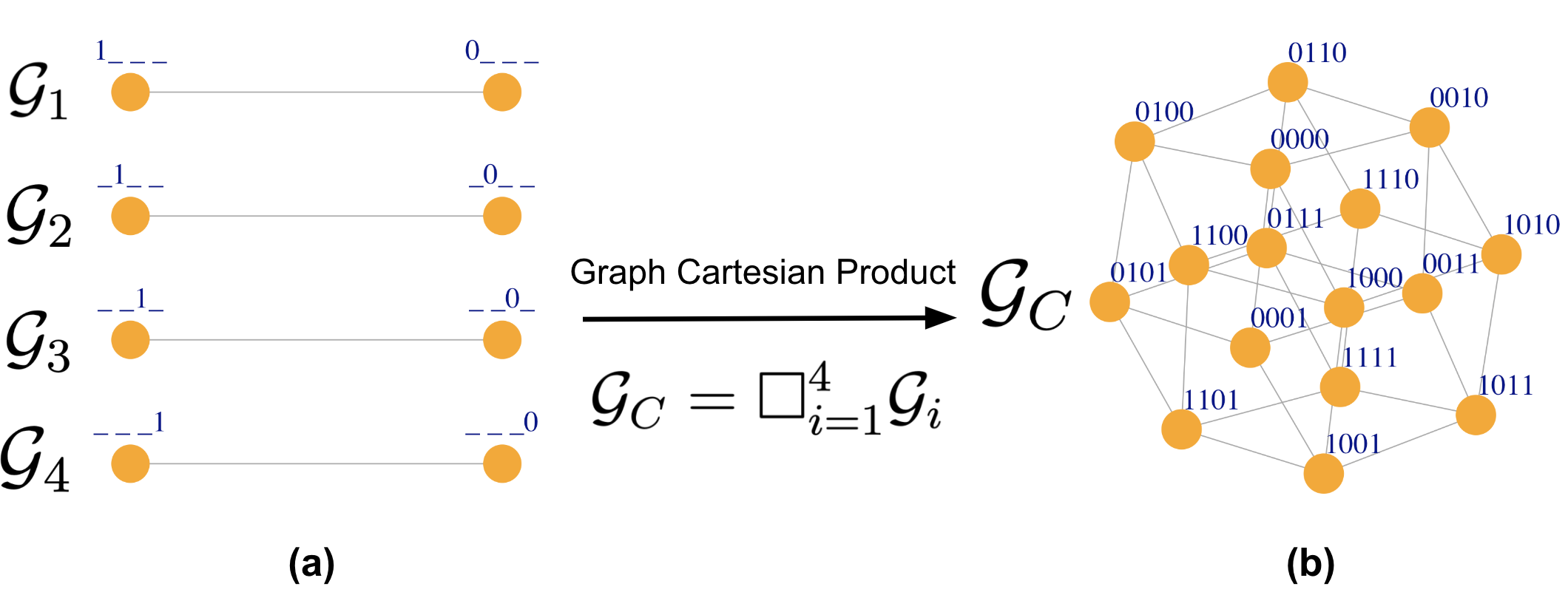}
        \caption{In (a), we have the sub-graphs $\mathcal G_i$ corresponding to each node $i$. In (b), we take the graph Cartesian product of these sub-graphs to obtain the combinatorial graph, $\mathcal G_C$.}
        \label{fig:combo2}
\end{figure}

Now, to construct the combinatorial graph in general, the authors propose to use the graph Cartesian product. Specific to our problem, let $\mathcal G_i$ be the sub-graph corresponding to node $i$ for $i=1,\dots,n=4$. Then $\mathcal G_i$ is a sub-graph with two vertices, where one vertex corresponds to node $i$ being included in the seed set, and the other vertex corresponding to node $i$ being excluded from the seed set, with an edge connecting these two vertices. We construct such a sub-graph for each node $i$, and then the combinatorial graph, $\mathcal G_C$ is the graph Cartesian product, i.e., $\mathcal G_C = \square_i \mathcal G_i$. If $\mathcal H_1$ and $\mathcal H_2$ are two graphs with vertices $\mathcal V_1$ and $\mathcal V_2$, respectively, then the graph Cartesian product is $\mathcal H=H_1\square H_2$ where $\mathcal H$ has vertices $\mathcal V_1\times\mathcal V_2$ and vertices $(h_1,h_2)$ and $(h_1',h_2')$ have an edge if and only if $h_i=h_i'$ and $h_j$, $h_j'$ have an edge in $\mathcal H_j$ for $i,j=1,2$.

In Figure \ref{fig:combo2}(a), we show the sub-graphs $\mathcal G_i$ where $i=1,2,\dots,4$. The vertex on the left-hand side corresponds to node $i$ being included in the seed set, while the right-hand side vertex corresponds to the node not being included in the seed set. In Figure \ref{fig:combo2}(b), we show the graph Cartesian product $\mathcal G=\square_{i=1}^4\mathcal G_i$. The difference between the graphs in Figure \ref{fig:combo1} and Figure \ref{fig:combo2}(b) is clear. Not only are the number of vertices and edges different, but the graph constructed using the graph Cartesian product has several vertices which correspond to inadmissible seed sets, e.g., $0100$ only seeds one node while $1101$ seeds three nodes. The discrepancy is because the graph Cartesian product does not preserve our cardinality constraint. 

While, technically, $\mathsf{COMBO}$ only requires constructing the combinatorial graph, all results in \cite{oh2019combo} directly rely on constructing this graph using the graph Cartesian product. Since we cannot construct the graph in this way for our specific problem, we are unable to use $\mathsf{COMBO}$ for the IM task. We consider it an interesting avenue of future work to extend the ideas of $\mathsf{COMBO}$ to constrained input spaces.

\section*{3. Greedy algorithm for $AEI(\cdot)$}

\begin{algorithm}[H]
\SetAlgoLined
\KwResult{Optimal seed set $\bx^*$}
 {\bf Input:} number of nodes $n$, seed set size $k$, acquisition function $AEI(\cdot)$ \;
 
 Initialize seed set $\bx^*$ such that $\sum_{i=1}^n x_i^*=k$\;
 
 $run = 1$\;
 
 \While{$run > 0$}{
 
 $run = 0$\;
 
 Randomly order nodes\;
 
  \For{$i$ such that $x_i^*=1$}{
    \For{$j$ such that $x_j^*=0$}{
        $\tilde \bx=\bx^*;\ \tilde x_i = 0;\ \tilde x_j = 1$\;

        \If{$AEI(\tilde \bx) > AEI(\bx^*)$}{
            $\bx^*=\tilde \bx$\;
            
            $run = 1$\;        
            }
    }
   
  }
 }
 \caption{Greedy}
\end{algorithm}

\section*{4. Connection with Trust regions}
The greedy algorithm can be seen as a relative of the Trust Region (TR) methodology, made popular in e.g., \cite{eriksson2019scalable, wan2021think, papenmeier2023bounce}. Let $\boldsymbol{x}^*$ be the current best solution, i.e., seed set which has yielded the largest influence spread. Then $\mbox{TR}_L(\boldsymbol{x}^*)$ is the Trust region of radius $L$ around $\boldsymbol{x}^*$, defined as all points within Hamming distance $L$ of $\boldsymbol{x}^*$, i.e.,
$$
    \mbox{TR}_L(\boldsymbol{x}^*)
    =\left\{\boldsymbol{x}\mid d_H(\boldsymbol{x}, \boldsymbol{x}^*)\leq L,\ \sum_{i=1}^n x_i=k\right\}.
$$
Of course for our problem, we can only consider points within the TR that also satisfy our cardinality constraint, which yields the second condition required for the TR.

Now, instead of searching the entire combinatorial input space for the optimal solution, a local search is performed within the TR. The radius $L$ is adjusted dynamically throughout the algorithm, increasing if a new optimum is found, and decreasing otherwise. When the radius falls below a certain threshold, the search may restart at a new initial condition.

With this in mind, we can see how the proposed greedy algorithm is similar to the TR framework. In the greedy algorithm, we have some current seed set $\boldsymbol{x}^*$ and we swap one seed node with an unseeded node to generate a candidate seed set $\tilde{\boldsymbol{x}}$. It is clear that
$$
    d_H(\boldsymbol{x}^*, \tilde{\boldsymbol{x}})
    =2,
$$
since the two seed sets exchange a single node.
Note that two is also the shortest possible distance two eligible seed sets can be; a Hamming distance of one is impossible as it would imply different sized seed sets. Thus, the greedy algorithm can be considered as a modified TR search centered on $\bx^*$ with $L=2$ fixed throughout the algorithm, albeit with a more restrictive search path.

\section*{5. IM for large networks}
\begin{figure}
    \centering
    \includegraphics[width=\textwidth]{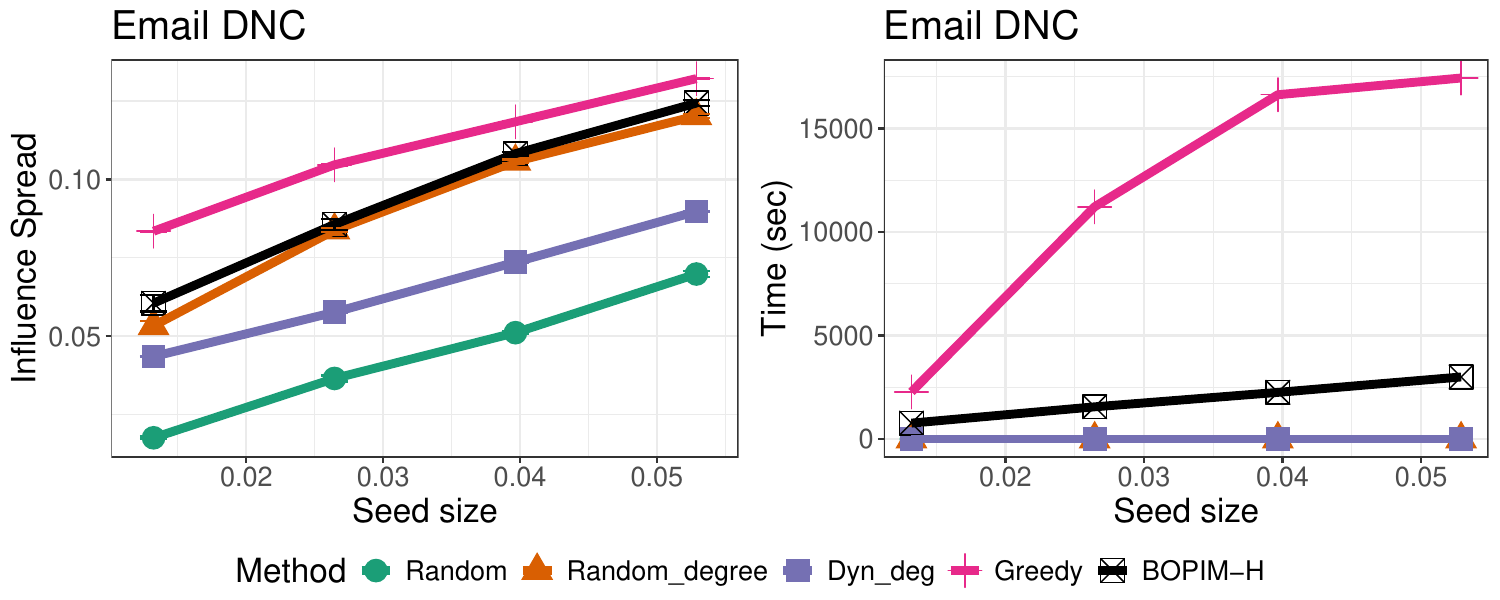}
    \caption{Results for Email DNC network.}
    \label{fig:large}
\end{figure}

To demonstrate the performance of $\mathsf{BOPIM}$ on large networks, we present the results for IM on the DNC email network \citep{nr} with $n=1891$ nodes. We fix $\lambda=0.05$ and $T=10$ while varying $k=25, 50,\dots, 100$. Because of the long run times, we only consider $\mathsf{BOPIM-H}$ using 5 MC repetitions, while still using 25 for Random and Random Degree. The influence spread results and run times are in Figure \ref{fig:large}. We can see that the proposed method provides comparable results to the greedy algorithm, but is significantly faster. While $\mathsf{BOPIM-H}$ performs similarly to Random Degree, it greatly outperforms Dynamic Degree.

\section*{6. Surrogate function validation}\label{sec:sim1}
\subsection*{Settings}
In these simulations, we quantify the fit of the surrogate model to the observed values of the objective function. Towards this end, we first run Algorithm 1 with $N_0=5$ and $B=20$ and save the final posterior distribution of $\beta_0$. Using this posterior distribution, we then compute out-of-sample influence spread predictions on $N=100$ seed sets and compare with the observed values of the objective function. If the surrogate model fits well, then its predicted influence spread will be close to the true influence spread. We quantify this similarity using the mean absolute prediction error (MAPE). If $\hat\beta_0$ is the posterior median of $\beta_0$, and $\boldsymbol{x}^*$ corresponds to the new seed set, then 
$$
    \tilde y
    = \hat\beta_0+\kappa(\boldsymbol{x}^*, {\bf X})^T(\gamma{\bf K}+{\bf I}_{N_0+b})^{-1}(\boldsymbol{y}-\hat\beta_0 {\bf 1}_{N_0+b})
$$
where $\kappa(\boldsymbol{x}^*, {\bf X})_j = \kappa(\bx^*,\bx_j)$ for $j=1,\dots,N_0+B$ and ${\bf 1}_n$ is the vector of ones of length $n$. The MAPE is computed as
$$
    \frac1{N}\sum_{i=1}^N |\tilde y_{i}-y_i|
$$
where $N$ is the number of out-of-sample seed sets. A smaller MAPE means a closer fit.

We stress that the goal of this experiment is to ensure that the surrogate model reasonably approximates the influence spreading process. Note that the observed values of the influence spread, $y_i$, are treated as the ground-truth, even though they were observed with noise from MC simulations. Thanks to the BO framework, we could obtain an entire predictive distribution, i.e. $p(\tilde y|{\bf X}, \boldsymbol{y})$. While quantifying the uncertainty in our predictions of the influence spread is a by-product of this algorithm, it is not the primary goal of this section. Rather, we simply want to know if the predictive mean is similar to the observed influence spreads. Thus, MAPE is an appropriate metric of interest.

We compute the MAPE for each of the four data sets. We fix $\lambda=0.05, 0.05, 0.01, 0.01$, and $k=3,4,6,10$ for the Reality, Hospital, Bluetooth and Conference 2 datasets, respectively. $k$ was chosen as approximately 5\% of the total number of nodes for each network. We consider the proposed algorithm using both the Hamming ($\mathsf{BOPIM-H}$), as well as the Jaccard kernel ($\mathsf{BOPIM-J}$). As a baseline method, we also consider intercept-only predictions, where we simply take the mean influence spread of 25 seed sets randomly sampled proportional to a node's degree.

For the out-of-sample predictions, we consider two node sampling schemes. One samples nodes at random while the other samples proportional to their degree on the aggregate network, as in $\mathsf{BOPIM}$. We expect the model to have a closer fit when the test points are sampled in the same way as the training points. Results are averaged over 25 Monte Carlo (MC) replications.

\subsection*{Results}
The results are in Table \ref{tab:val}. Note that the MAPE results are interpreted as follows: for the Hamming kernel on the Conference 2 network with degree-based sampling of seed sets, for example, the MAPE value of 1.06 means that, on average, the estimated influence spread of the surrogate function is within about one node of the true influence spread. We can see that for all networks and both sampling schemes, the MAPE values are quite small. We also notice very little difference in the results between the two kernels and intercept-only model. This is not too surprising as the $\mathsf{BOPIM}$ algorithm also uses an intercept-only mean function, and here we are making predictions about the mean. Of course, the intercept-only predictions cannot be used for seed selection as it gives no information on node importance.

Additionally, the degree-based sampling always leads to smaller MAPE values, which is to be expected since this is how the $\mathsf{BOPIM}$ algorithm samples points for the initial fitting stage. Indeed, the surrogate model fits the true objective function more closely when the seed nodes have a large degree, which is where we hypothesize that the global optimum lies. These results give use confidence that our surrogate function is doing an adequate job of modeling the true objective function. 

\begin{figure}
    \centering
    \includegraphics[width=\linewidth]{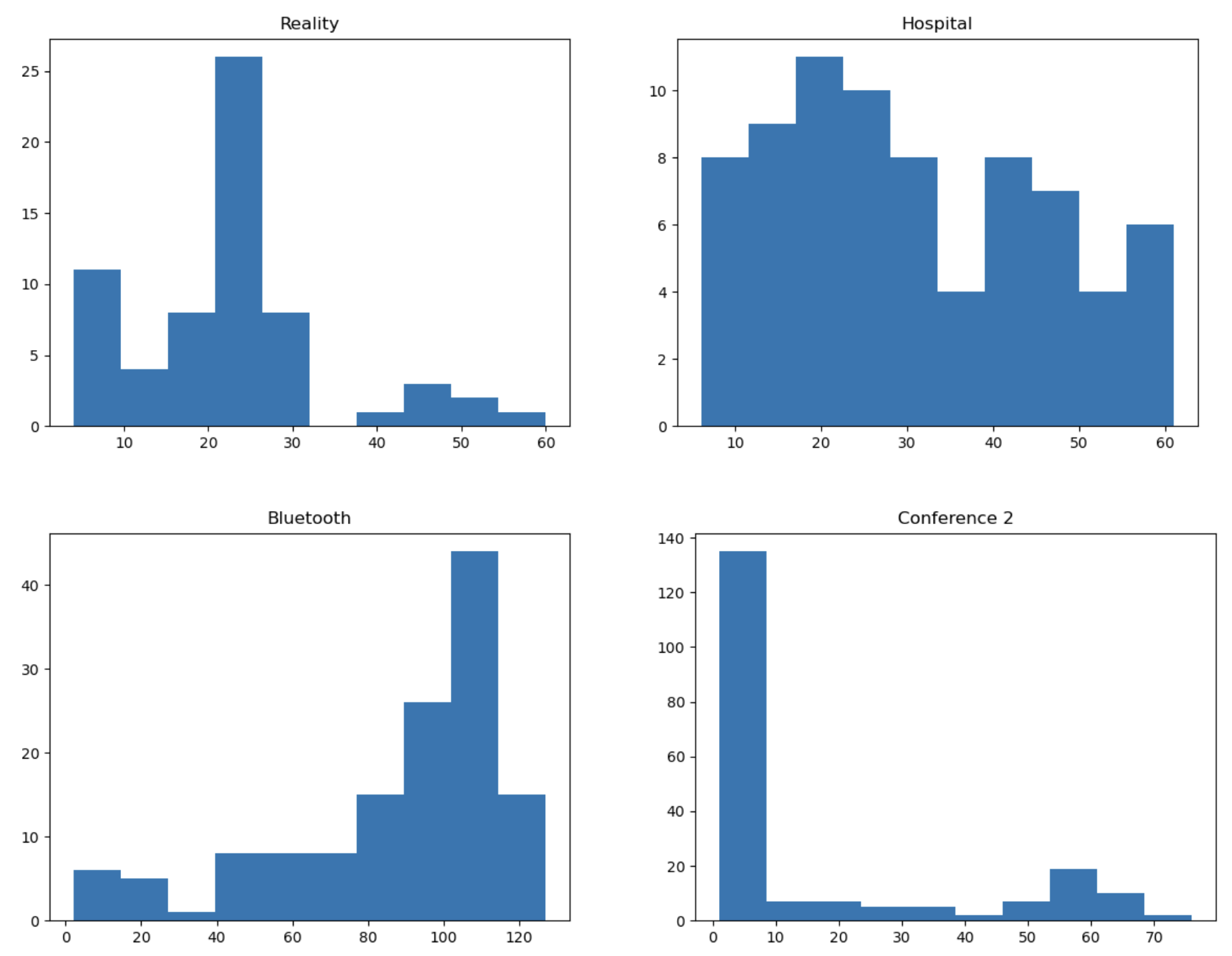}
    \caption{Degree distributions for the networks in this paper.}
    \label{fig:deg}
\end{figure}

Finally, while for all networks the Degree setting yields lower MAPE values, this difference is most pronounced for the Conference 2 network. This is likely due to the degree distributions of the various networks. In Figure \ref{fig:deg}, we report a histogram of the degree distribution for each network. For Conference 2, the vast majority of nodes have a very low degree. Since there are fewer nodes with large degree, when we sample proportionally to the node's degree, it is more likely that the same nodes are being chosen. This increases the likelihood that the out-of-sample seed sets are similar to the seed sets that the model was already trained on, lowering the MAPE value. On the other hand, when the out-of-sample seed sets are randomly sampled, we will sample many nodes with very low degree, yielding seed sets unlike those that the model has been trained on, increasing the MAPE values.

\begin{table*}[]
    \centering
    \begin{tabular}{lll|c}
        Dataset & Sampling & Kernel & MAPE (se) \\\hline
        Reality & Random & Intercept & 3.77 (0.08)  \\
                &        & Hamming & 4.24 (0.23)\\
                &        & Jaccard & 4.65 (0.22)\\
                & Degree & Intercept & 3.42 (0.06)\\
                &        & Hamming & 3.09 (0.14)\\
                &        & Jaccard & 3.37 (0.13)\\\hline
        Hospital & Random & Intercept & 4.76 (0.09)  \\
                &        & Hamming & 4.58 (0.16)\\
                &         & Jaccard & 4.61 (0.12)\\
                & Degree  & Intercept &  3.05 (0.05) \\
                &         & Hamming & 2.92 (0.07)\\
                &         & Jaccard & 2.86 (0.08)\\\hline
        Bluetooth & Random & Intercept & 2.49 (0.07) \\
                &        & Hamming & 2.62 (0.09) \\
                &          & Jaccard & 3.40 (0.16) \\
                & Degree & Intercept & 1.59 (0.02) \\
                &        & Hamming & 1.56 (0.05) \\
                &         & Jaccard & 2.02 (0.09)\\\hline
        Conference 2 & Random & Intercept & 7.86 (0.07)  \\
                     &        & Hamming & 6.87 (0.13)\\
                     &        & Jaccard & 5.06 (0.10)\\
                     & Degree & Intercept & 1.01 (0.02)  \\ 
                     &        & Hamming & 1.06 (0.04)\\
                     &        & Jaccard & 1.83 (0.11)
    \end{tabular}
    \caption{Mean absolute prediction errors for validating surrogate function simulations. Standard error in parentheses.}
    \label{tab:val}
\end{table*}

\section*{7. Initial sampling ablation study}
In this section, we consider an alternative sampling scheme for the initial observations in the BO algorithm. We remark that nodes with high degrees on the temporally aggregated network, $\tilde G=\cup_t G_t$, are good candidates for the optimal seed set. Additionally, we would like to sample nodes that are ``far apart'' on the graph to ensure better influence spread. Therefore, instead of randomly sampling $\bx_i$ from $\mathcal X$, we sample nodes proportional to a metric which accounts for degree and distance from the other nodes in the seed set. For the first node, we sample it proportional to the degree on the temporally aggregated network. Mathematically, this means sampling from a distribution with probability mass function $\mathsf{P}(Z=j)=d_j/\sum_k d_k$ where $d_j$ is the degree of node $j$ on $\tilde G$ for $Z\in\{1,\dots,n\}$. For the remaining $K-1$ nodes, we compute the average distance between the node $i$, and nodes already sampled for the seed set, call it $\nu_i^{(S)}$. We standardize both of these measures by converting to a $z$-score, i.e.,
    $$
        \tilde d_i=\frac{d_i-\bar d}{s_d},\ \ \tilde\nu_i=\frac{\nu_i-\bar\nu}{s_\nu}
    $$
    where $\bar d$ and $s_d$ are the sample mean and standard deviation of the degrees, $d_1,\dots,d_n$, respectively, and similarly for $\bar\nu$ and $s_\nu$. Then we sample nodes proportionally to $\boldsymbol{p}_\varrho = (p_1,\dots,p_n)$
    where
    $$
        p_i
        = \tilde d_i + \varrho \tilde \nu_i
    $$
    where $\varrho>0$ is a tuning parameter which controls between preferring nodes with high degree (small $\varrho$) with preferring nodes far apart (large $\varrho$). For all experiments, we set $\varrho=1$ for an equal balance. This sampling scheme ensures that the initial seed sets prioritize large degree nodes that are not nearby on the graph.

    We compare this new sampling scheme (coined Space-degree) with the originally proposed sampling scheme (Degree-only) on the Reality network for $k=2,3,\dots,7$. For each value of $k$, we run the $\mathsf{BOPIM-H}$ algorithm, obtain the optimal seed set and compute the influence spread. This process is repeated 25 times and the average influence spread is reported in Table \ref{tab:int_ab}. The influence spreads of the two methods are almost identical, so we only present the simpler Degree-only sampling scheme in the main manuscript.

    \begin{table}[]
        \centering
        \begin{tabular}{c|ccccccc}
            Sampler / $k$ & 2 & 3 & 4 & 5 & 6 & 7  \\\hline
            Degree-only & 22.0 (0.2) & 25.1 (0.1)& 27.4 (0.1) & 29.2 (0.1) & 30.5 (0.1) & 31.7 (0.1) \\
            Space-degree & 21.9 (0.3)& 25.2 (0.1) & 27.3 (0.2) & 29.2 (0.1) & 30.8 (0.1) & 31.6 (0.1)
        \end{tabular}
        \caption{Average influence spread (with standard error) for the degree-only and space-degree initial sampling schemes.}
        \label{tab:int_ab}
    \end{table}

\section*{8. Robustness experiments}

In this section, we explore the robustness of the proposed method to varying the size of the seed set as well as the number of temporal snapshots.

\subsection*{Varying $k$}
First, we are interested in the performance of the proposed method when the fitting steps are done on one value of $k'$, but then we find the optimal seed set for $k\neq k'$. Specifically, we let $k'$ be fixed and run $\mathsf{BOPIM-J}$ for $B$ BO loop iterations as usual. After the last iteration, however, we re-fit the GP regression model and use the acquisition function to find the optimal seed set of size $k$ which is not necessarily the same as $k'$. We then use this seed set to compute the number of influenced nodes on the network.

In the following experiments, we fix $\lambda=0.05$ and $T=10$ and look at the Reality and Hospital networks. All other hyper-parameters are the same as in the previous experiments. We consider four different methods, Low, Mid, High and Mix. For Low, Mid and High, the BO routine is carried out setting $k'=2, 4$ and $7$, respectively. For Mix, we randomly sample $k'$ from a uniform distribution on $\{2,3,\dots, 7\}$ at each stage in the initial fitting as well as the BO loop. For each method, we then compute the optimal seed set of size $k$ for $k=2,3,\dots, 7$. The results are repeated 25 times and averaged for each value of $k$. We compare the results with the original $\mathsf{BOPIM}$ algorithm where the same $k$ is used in the training phase of the algorithm as well as the final influence spread evaluation.

Note that even when the network is trained on seed set of size $k'$ and then tested on sets of size $k=k'$, this is still not equivalent to the original $\mathsf{BOPIM}$ algorithm. In the original algorithm, the model is fit on one value of $k$ throughout the $B$ BO iterations and then the set seed which corresponded to the largest influence spread is then chosen as the optimal set. In these robustness simulations, on the other hand, the final posterior distribution is used to select the optimal seed set. Therefore, Mid, for example, is not equivalent to $\mathsf{BOPIM}$ even when $k'=k=4$.

\begin{figure}
    \centering
    \includegraphics[width=\textwidth]{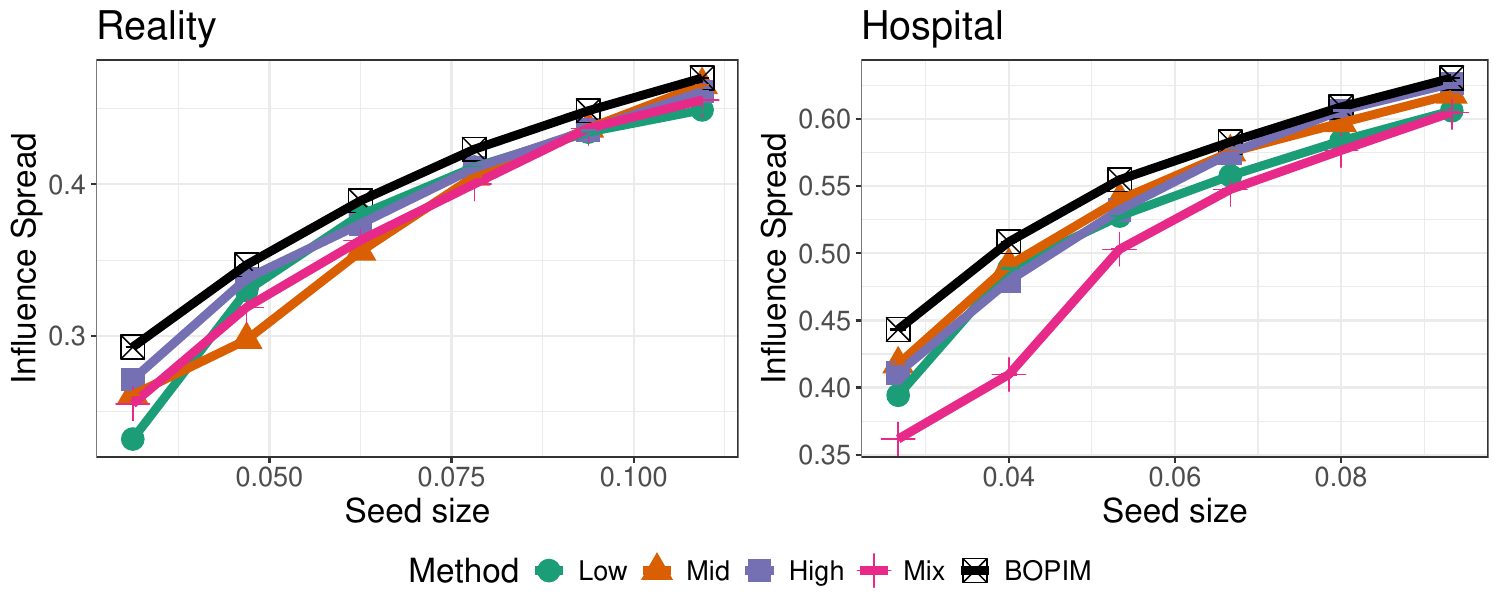}
    \caption{Robustness simulations where the size of the seed set in the model fitting is different from the final seed set.}
    \label{fig:k_train}
\end{figure}

The results are in Figure \ref{fig:k_train}. In the Reality network, High tends to consistently comes closest to the original $\mathsf{BOPIM}$ algorithm for all $k$. The relative performance of Low decreases as $k$ increases (for $k>2$), while Mid's relative performance improves. Mix, however, yields seed sets with noticeably smaller influence spreads. The trends are similar in the Hospital network, where High and Mid consistently yields influence spreads closest to that of the original $\mathsf{BOPIM}$ algorithm, while Low's performance is best for $k=3,4$. The performance of Mix is even worse relative to the other methods compared to the Reality network.

These findings are promising for the performance of the $\mathsf{BOPIM}$ algorithm. In particular, it means that even if we run the algorithm for one value of $k'$, the results can be used to select seed sets of various sizes. Additionally and not surprisingly, the general trends show that as $|k'-k|$ increases, the performance seems to decrease slightly. Thus, the method will yield better seed sets when $k'\approx k$. This not only shows the generalizability of the method, but also can save computing time as the algorithm may not need to be re-run for each value of $k$. Finally, a single $k'$ should be used in the algorithm's model fitting as the Mix method consistently performed poorly.

\subsection*{Varying $T$}
This sub-section is similar to the previous, except now we are looking at the robustness of the $\mathsf{BOPIM}$ algorithm to different values of $T$, the number of temporal snapshots. In particular, we run the $\mathsf{BOPIM}$ algorithm for some fixed $T'$ and obtain the optimal seed set. This seed set is then used to compute the influence spread on the same network but now with $T\in\{2,3,\dots,10\}$ snapshots. This scenario can be considered as model mis-specification as the number of snapshots is not correctly known.

We fix $\lambda=0.05$ and $k=3,4$ for the Reality and Hospital network, respectively. We now consider three different methods, Low, Mid and High. For these methods, we train the proposed method on the network using $T'=2,5$ and $10$, respectively. We then use the resulting seed sets to compute the influence spread on the same network but with $T=2,3,\dots,10$. The results are again averaged over 25 repetitions for each value of $T$ and compare with the $\mathsf{BOPIM}$ algorithm fit using the correct value of $T$.

\begin{figure}
    \centering
    \includegraphics[width=\textwidth]{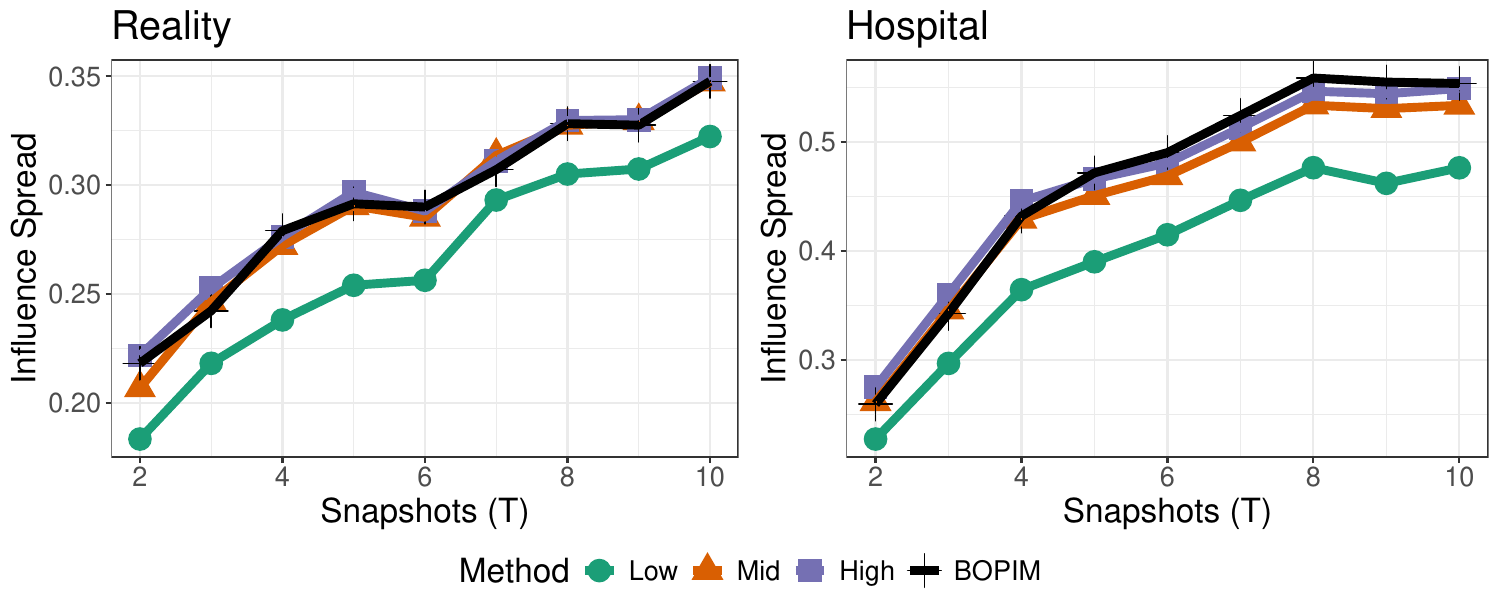}
    \caption{Robustness simulations where the number of temporal snap shots in the model fitting is different from influence spread calculation.}
    \label{fig:T_train}
\end{figure}

The results are in Figure \ref{fig:T_train}. For the both networks, Mid and High perform similarly and almost perfectly approximate the correctly-specified algorithm. Low, however, yields seed sets with much smaller influence spreads. Again, these results are favorable for the $\mathsf{BOPIM}$ algorithm. They show that the algorithm can be fit on almost any value of $T'$ and the resulting seed set will yield large influence spreads for nearly all $T$.  Moreover, they imply that the optimal seed set is not very sensitive to the number of snapshots used to aggregate the temporal network. For example, if a node is influential with $T=5$ snapshots, it is also likely to be influential with $T=10$ snapshots. This is somewhat expected as central nodes should be influential regardless of the network aggregation scheme. Thus, the proposed algorithm is relatively robust to the choice of $T$. 

\section*{9. Horseshoe prior Gibbs sampler}
In this section, we report the Horseshoe prior \citep{polson2010shrink} Gibbs sampler, using the auxiliary variable formulation of \cite{makalic2015simple}.
\begin{enumerate}
    \item Sample $\bbeta|\boldsymbol{\lambda},\boldsymbol{\nu}, \tau^2,\sigma^2, {\bf Y}\sim\mathsf{N}(\boldsymbol{\mu},\sigma^2{\bf V})$ where $\boldsymbol{\mu}={\bf V}{\bf X}^T{\bf Y}$, ${\bf V}=({\bf X}^T(\gamma{\bf K}+{\bf I})^{-1}{\bf X}+{\bf S}^{-1})^{-1}$, \\
    ${\bf S}= diag(\eta_1^2\tau^2,\dots, \eta_n^2\tau^2)$.
    \item Sample $\sigma^2|\bbeta,\boldsymbol{\eta},\boldsymbol{\nu}, \tau^2, {\bf Y}\sim\mathsf{IG}(a_1+(N_0+n)/2, b_1+(({\bf Y}-{\bf X}\bbeta)^T(\gamma{\bf K}+{\bf I})^{-1}({\bf Y}-{\bf X}\bbeta)+\bbeta^T{\bf S}^{-1}\bbeta)/2)$.
    \item Sample $\eta_j^2|\bbeta,\boldsymbol{\nu},\tau^2,\sigma^2\sim\mathsf{IG}(1, \nu_j^{-1}+\beta^2_j/(2\tau^2\sigma^2)$, $j=1,\dots,n$
    \item Sample $\tau^2|\bbeta,\boldsymbol{\eta}, \sigma^2,\xi\sim\mathsf{IG}((n+1)/2, \xi^{-1}+\sum_{k=1}^n \beta^2_k/(2\eta^2_k\sigma^2))$.
    \item Sample $\nu_j|\eta_j\sim\mathsf{IG}(1,1+\eta_j^{-2})$ for $j=1,\dots,n$.
    \item Sample $\xi|\tau^2\sim\mathsf{IG}(1, 1+\tau^{-2})$.
\end{enumerate}

\end{document}